\documentclass[12pt, draftclsnofoot, onecolumn]{IEEEtran}
\usepackage[cmex10]{amsmath}
\usepackage{amssymb}
\usepackage{cite}
\usepackage{graphicx}
\usepackage{array,color}
\usepackage{multirow}
\usepackage{amsmath}
\usepackage{stfloats}
\usepackage{graphicx}
\usepackage{subfigure}
\usepackage{tabularx}
\usepackage{epsfig,epsf,color,balance,cite}
\usepackage{setspace}
\usepackage{bm}
\usepackage{textcomp}
\usepackage{algorithmic}
\usepackage{algorithm}
\usepackage{subfigure}
\usepackage{caption}
\usepackage{graphicx}
\usepackage{setspace}

\begin{document}

\title{Joint Pilot and Payload Power Allocation for Massive-MIMO-enabled URLLC IIoT Networks}

\author{Hong Ren, Cunhua Pan, Yansha Deng, Maged Elkashlan, Arumugam Nallanathan, \IEEEmembership{Fellow, IEEE}

\thanks{H. Ren, C. Pan, M. Elkashlan and A. Nallanathan are with School of Electronic Engineering and Computer Science, Queen Mary University of London, London, E1 4NS, U.K. (Email:{h.ren, c.pan, a.nallanathan}@qmul.ac.uk). Y. Deng is with the Department of Informatics, King's College London, London WC2R 2LS, U.K. (e-mail:yansha.deng@kcl.ac.uk). }
}

\maketitle

\vspace{-0.4cm}
\begin{abstract}
The Fourth Industrial Revolution (Industrial 4.0) is coming, and this revolution will fundamentally enhance the way the factories manufacture products. The conventional wired lines connecting central controller to  robots or actuators will be replaced by wireless communication networks  due to its low cost of maintenance and high deployment flexibility. However, some critical industrial applications require ultra-high reliability and low latency communication (URLLC). In this paper, we advocate the adoption of massive multiple-input multiple output (MIMO) to support the wireless transmission for industrial applications as it can provide deterministic communications similar as wired lines thanks to its channel hardening effects. To reduce the latency, the channel blocklength for packet transmission is finite, and suffers from transmission rate degradation and decoding error probability. Thus, conventional resource allocation for massive MIMO transmission based on Shannon capacity assuming the infinite channel blocklength is no longer optimal. We first derive the closed-form expression of lower bound (LB) of achievable uplink data rate for massive MIMO system with imperfect channel state information (CSI) for both maximum-ratio combining (MRC) and zero-forcing (ZF) receivers. Then, we propose novel low-complexity algorithms to solve the achievable data rate maximization problems by jointly optimizing the pilot and payload transmission power for both MRC and ZF. Simulation results confirm the rapid convergence speed and performance advantage over the existing benchmark algorithms.
\end{abstract}


\IEEEpeerreviewmaketitle
%
\vspace{-0.3cm}
\section{Introduction}
Industry 4.0 has been envisioned as the future paradigm for the next generation of industrial systems, which integrates advanced manufacturing functions with the industrial internet-of-things (IIoT) to create a more intelligent and automatic digital manufacturing system \cite{drath2014industrie}. Traditionally, industrial control systems mainly rely on wired connections such as cables or optical fiber, since the current wireless networks cannot meet their stringent latency and reliability requirements. However, there are some drawbacks to deploying wired lines. First, significant cost will be incurred by the installation and maintenance. Second, wired lines are vulnerable to wear and tear in motion control applications, and suffer from aging. Finally, they cannot be deployed in some harsh environments, such as those with high temperatures and rotating part. Hence, to make Industry 4.0 a reality, it is imperative to design wireless networks tailored for industrial applications to replace the traditional wired lines. Typical industrial applications require deterministic communications with  ultra reliability ($1-10^{-9}$) and  low latency ($1$ ms), such as factory automation (FA) \cite{hongtwc2019}, power system protection (PSP), and power electronics control (PEC) \cite{pang2017wireless}. Significant research efforts have been devoted to the design of wireless communications in industrial applications. However, most of the existing works mainly focused on the adaption of the upper layers of conventional wireless networks to achieve deterministic communications, while keeping the physical layer untouched. Some related standards are WirelessHART, wireless interface
for sensors and actuators (WISA), and Wireless Networks for Industrial Automation/Process Automation (WIA-PA) \cite{luvisotto2016ultra}.  Although keeping the wireless standards of physical layers can allow faster design and better compatibility, it leads to a fundamental bottleneck for the system performance. None of the above standards can meet the stringent demand requested by the most critical FA, PSP and PEC applications. As a result, more efforts should be devoted to the design from the physical layer perspective of view.

From the physical layer perspective, the dominate feature of the industrial applications is that the packet transmission should be completed within short blocklength due to low latency requirement \cite{ma2019high}. Hence, the transmission is not error-free with any finite/short blocklength channel codes. In this case, Shannon capacity formula is not applicable since it is based on the principles of the law of large numbers, and we need to design the resource allocation by considering the decoding error probability requirement. In \cite{Polyanskiy2010}, Peter \emph{et al.}  have derived the approximation formula of the maximum achievable data rate with finite blocklength transmission, which characterises the complicated relationships among decoding error probability, channel blocklength, and signal-to-noise ratio (SNR). Unlike Shannon capacity formula, the data rate expression under short packet transmission is neither convex nor concave with respect to the SNR or blocklength \cite{she2017radio}. As a result, the optimal resource allocation under this formula is difficult to obtain.

Recently, there are increasing research studying the transmission design based on the short packet transmission capacity formula \cite{xiaoyusun,yulinhu2018,pan2019joint,hongicc2019,chen2019resource,changyang2018,ghanem2019resource}. In specific, the effective throughput maximization was studied in \cite{xiaoyusun} for a two-device downlink non-orthogonal multiple access (NOMA) system. The overall error probability is minimized in \cite{yulinhu2018} for a simultaneous wireless information
and power transfer (SWIPT)-enabled decode-and-forward (DF) relaying network. The decoding error probability minimization was investigated in \cite{pan2019joint} for a unmanned aerial vehicle (UAV)-enabled DF relay system. We recently proposed a low-complexity power  and blocklength optimization algorithm  for both orthogonal multiple access (OMA) and NOMA in a two-hop relay system in \cite{hongicc2019}. However, all these contributions are limited to a simple scenario with two devices, where one device acts as a relay and the other as the destination node. In industrial applications, the central controller needs to support a large number of devices \cite{luvisotto2016ultra}. On one hand, Chen \emph{et al.}  in \cite{chen2019resource} investigated the effective capacity maximization problem for wireless-powered IoT network with multiple devices operating in a time division multiple access (TDMA) mode. However, the time budget is already tight, and the portion allocated to each device will be marginal. Hence, TDMA strategy is not suitable for the applications in industrial applications with extremely stringent latency target. On the other hand, the authors in \cite{changyang2018} \cite{ghanem2019resource} studied the resource allocation for multiple devices operating under the orthogonal frequency division multiple access (OFDMA) mode. Unfortunately, this requires huge amount of system bandwidth, which is not feasible as some industrial applications operate over unlicensed spectrum \cite{huang2018new}.

By equipping a large number of antennas at the base station (BS), massive multiple-input multiple-output (MIMO) has been widely regarded as the key enabler for the creation of the fifth generation (5G) wireless networks \cite{andrews2014will}. By exploiting excessive number of spatial degrees of freedom, massive MIMO is capable of supporting multiple devices simultaneously without additional time or frequency resources. In addition, due to the channel hardening effect, massive MIMO  is more immune to the fast fading and can provide deterministic communications required by the industrial applications. Due to these attractive advantages, massive MIMO is ideal for supporting industrial applications with stringent quality of services (QoS) requirements. However, most of the existing literature adopted Shannon capacity as the performance metric to optimize the resource allocation \cite{cheng2017optimal,saatlou2019joint,lu2015joint},  which implicitly assumes the infinite channel blocklength. Therefore, conventional resource allocation solution based on Shannon capacity is not optimal for industrial applications with short channel blocklength. To the best of our knowledge, we are the first to study the resource allocation for massive MIMO providing ultra-reliability and low-latency communications (URLLC) for any number of devices.  Specifically, our contributions are summarized as follows:
\begin{enumerate}
  \item We derive the closed-form lower bounds (LBs) on the achievable rates for a  uplink massive MIMO system by considering the imperfect channel state information (CSI) with  finite channel blocklength, and assuming both maximum-ratio combining (MRC) and zero-forcing (ZF) receivers. They can be regarded as the conventional Shannon capacity minus a penalty term due to short packet transmission. Simulation results confirm the tightness of the LBs. Given fixed delay budget, we formulate an optimization problem with the objective to maximize the weighted sum rate by jointly optimizing the pilot and payload transmission power subject to the decoding error probability, the minimum data rates, and the energy constraints for all URLLC devices.
  \item For the case with MRC receiver, the formulated optimization problem is non-convex due to the complicated expression of data rate LBs, and it is difficult to find the globally optimal solution. To deal with this issue, we first approximate the penalty term in the data LB as a log-function, which can facilitate the transformation from the original optimization problem to a series of geometric programs (GPs). Each GP problem can be  efficiently solved with polynomial time. Besides, we provide a novel method to check the feasibility of the original problem, and provide both complexity and convergence analysis.
  \item For the case with ZF receiver, the data rate LB is more complicated and the algorithm proposed for MRC cannot be directly applied since the nominator of the signal-to-interference-plus-noise (SINR) is  a posynomial function. To handle this issue, we approximate  the polynomial functions with their best local monomial approximations and then transform the optimization into a series of GPs with low-complexity. Convergence analysis tailored for the ZF receiver is further provided.
  \item Simulation results show that our proposed algorithms converge rapidly, which verifies the low-complexity of the proposed algorithm. In addition, it is also shown that the proposed algorithm outperforms the benchmark schemes, especially the ones adopting Shannon capacity as the optimization performance metric, which emphasizes the importance of using short packet transmission theory.
\end{enumerate}

The remainder of this paper is organized as follows. In Section \ref{systemmodel}, system model and problem formulation are provided. In Section \ref{MRCcase},  we provide a low-complexity algorithm for joint design of pilot and payload power allocation for the MRC case. The ZF case is studied in Section  \ref{ZFcase}. Then, simulation results and analysis are presented in Section \ref{simulation}. The final conclusion is drawn in Section \ref{conclusion}.

\vspace{-0.3cm}
\section{System Model and Problem Formulation}\label{systemmodel}
\subsection{Factory System Model}

\begin{figure}[h]
\centering
\includegraphics[width=3.5in]{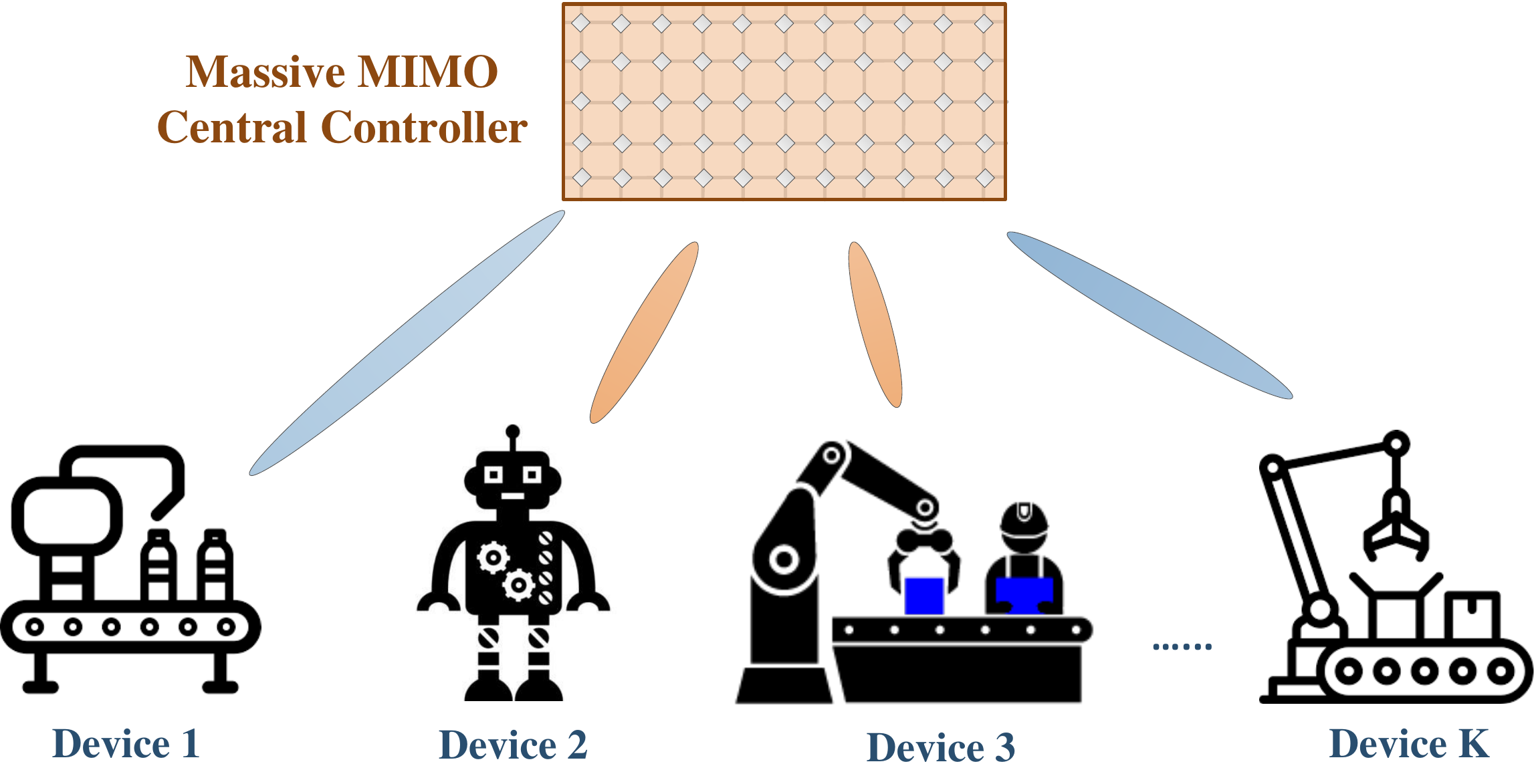}
\vspace{-0.2cm}
\caption{Factory scenario where a massive MIMO central controller serves multiple devices}
\label{systemmodelfig}
\vspace{-0.8cm}
\end{figure}

Consider a uplink multi-device Massive MIMO communication in one factory as shown in Fig. \ref{systemmodelfig}, where the central controller (CC) serves $K$ devices, e.g., actuator, robot, etc. The devices need to send their emergency information of URLLC requirements such as measured data or their current operation states to the CC. Thus, the CC can process these data information immediately and provide prompt response/feedback. For simplicity, we focus on the uplink transmission and the solutions for downlink can be similarly derived. The set of the devices is denoted as $\mathcal{K}=\{1,2,\cdots,K\}$. The CC is equipped with $M$ antennas and each device is equipped with single antenna due to their low signal processing capability, where $M \gg K$. Let us denote $\mathbf{h}_k \in \mathbb{C}^{M \times 1}$ as the channel vector from the CC to the $k$-th device and can be decomposed as $\textbf{h}_k=\sqrt{\alpha_k}\bar{\textbf{h}}_k$, where $\alpha_k$ denotes the large-scale channel gain that includes the pathloss and shadowing, and $\bar{\textbf{h}}_k$ denotes the small-scale fading following the distribution of $\mathcal{CN}(\textbf{0},\textbf{I})$. Let us denote $\textbf{H} \in \mathbb{C}^{M\times K}$ as the channel matrix from the $K$ devices to the CC, with $\textbf{H}=[\textbf{h}_1,\textbf{h}_2,\cdots,\textbf{h}_K]$.

The $K$ devices need  to transmit $K$ packets to the controller. Then, the $M \times 1$ received signal vector at the CC is given by
\vspace{-0.15cm}
\begin{equation}\label{jogihtioh}
\vspace{-0.15cm}
  {\bf{y}} = \sum\nolimits_{k \in \mathcal{K}} {{{\bf{h}}_k}\sqrt {p_k^d} } {s_k} + {\bf{n}},
\end{equation}
\!\!where $p_k^d$ is the payload power of the $k$th device, $s_k$ is the zero mean and unit variance Gaussian information message from the $k$th device, and ${\bf{n}} \sim \mathcal{CN}({\bf{0}}, {\bf{I}}_M)$ is the additive noise during the data transmission, where the variance of each element is normalized to unit.

\vspace{-0.3cm}

\subsection{Channel Estimation in Massive MIMO URLLC}\label{framestructure}
Due to the channel hardening \cite{Thomas} brought by massive MIMO, the system is more immune to the fast fading, which can provide high reliable services for the devices. However, to reap the benefits brought by massive MIMO, the CSI should  be available at the CC. Furthermore, TDD mode is always taken as an enabler for massive MIMO systems since downlink instantaneous CSI is obtained by estimating uplink CSI based on channel reciprocity \cite{KBmassiveTDD}. In addition, for machine type communications,
the devices may not be able to perform complicated signal processing tasks required in frequency division duplexing (FDD) systems, such as channel estimation calculation, quantization, etc. More time slots are needed for CSI feedback.
Hence, the TDD protocol is adopted in this paper.  All the devices should be allocated with orthogonal pilot resources so that the CC is able to distinguish the channels from different devices, thus, the number of symbols for
channel estimation should be no smaller than the number of devices\cite{Thomas}.

\begin{figure}[h]
\centering
\includegraphics[width=3.8in]{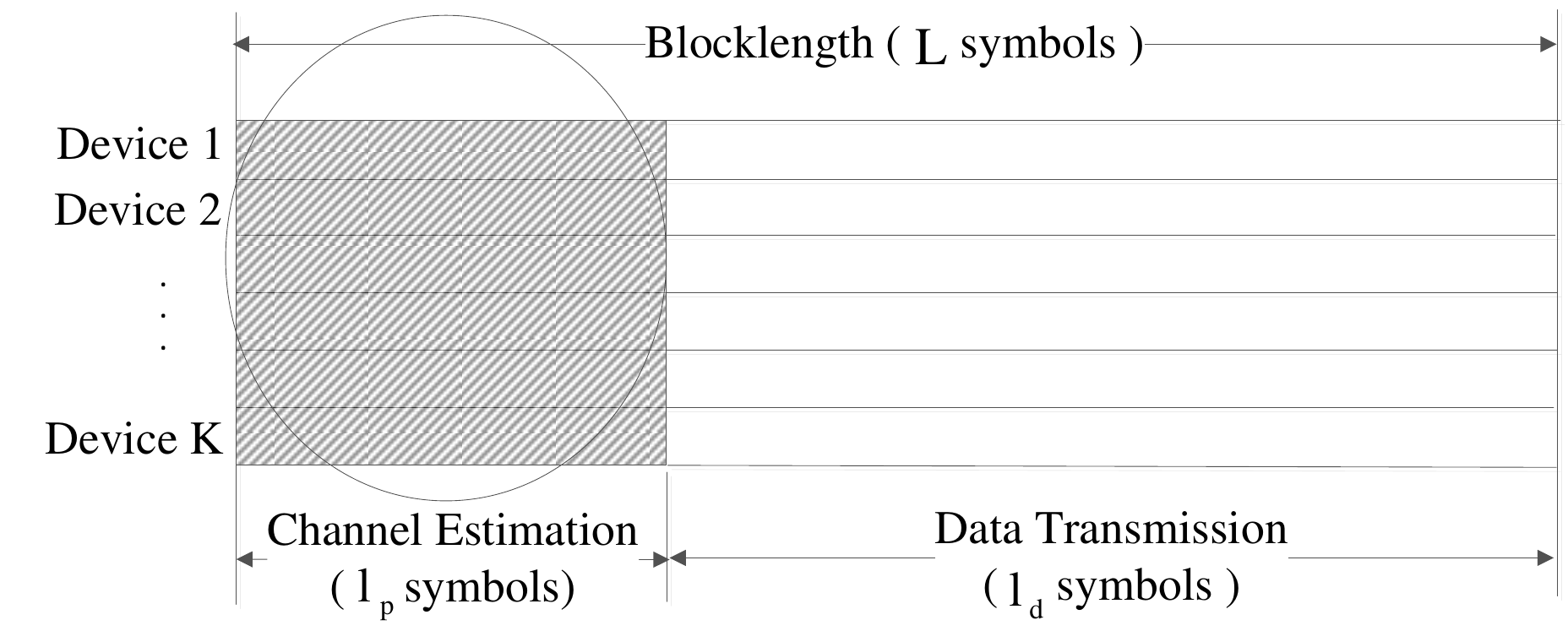}
\vspace{-0.2cm}
\caption{Block diagram scheme in TDD URLLC Massive MIMO scheme}
\label{Blockdiagram}
\vspace{-0.6cm}
\end{figure}

In a massive MIMO URLLC scenario, each block mainly consists of two parts: 1) $l_p$ symbols for channel estimation (the pilot sequence is of length $l_p$); 2) $l_d$ symbols for the $K$ devices' data transmission, thus the total number of symbols of the frame is denoted as $L=l_p+l_d$. As the symbols used for data transmission is already limited, we assume that $K$ devices sharing the same symbol duration and the frame structure is illustrated in Fig. \ref{Blockdiagram}.  Accordingly, the time durations for channel estimation and data transmission in one frame are given by $t_p=l_p/B$  and $t_d=l_d/B$, respectively, where $B$  is the bandwidth of the system.

In the training phase, all devices simultaneously and synchronously transmit orthogonal pilot sequences $\textbf{q}_1,\cdots, {\textbf{q}}_{l_p} \in {\mathbb{C}}^{l_p \times 1}$ to the CC, with ${\textbf{q}_k^H}\textbf{q}_k=1$ and ${\textbf{q}_i^H}\textbf{q}_j=0$, $i\neq j$. Hence, the minimum length of the pilot sequences to guarantee the orthogonality is equal to $l_p=K$. Based on the received signal, the CC estimates the channel conditions of all devices, and the received pilot signal at the CC is
\vspace{-0.2cm}
\begin{equation}
\vspace{-0.2cm}
\mathbf{Y}^p=\sum\nolimits_{k\in \mathcal{K}}\sqrt{Kp_k^p}\mathbf{h}_k\mathbf{q}_k^H+\mathbf{N},
\end{equation}
\!\!where $p_k^p$ is the pilot transmit power at the $k$th device, and $\mathbf{N}\in \mathbb{C}^{M\times K}$ is the additive Gaussian noise matrix received during the training phase, whose elements are independently generated and follow the distribution of $\mathcal{CN}(0, 1)$. To obtain channel $\textbf{h}_k$, the CC first multiplies $\textbf{Y}^p$ by $\frac{1}{\sqrt{Kp_k^p}}\textbf{q}_k$, which yields
\vspace{-0.2cm}
\begin{equation}
\vspace{-0.2cm}
\mathbf{y}_k^p=\frac{1}{\sqrt{Kp_k^p}}\textbf{Y}^p\textbf{q}_k=\textbf{h}_k+{\textbf{n}}_k^p,
\end{equation}
where ${\textbf{n}}_k^p=\frac{1}{\sqrt{Kp_k^p}}\textbf{N}\textbf{q}_k$. Since $\textbf{q}_k$ is a unit-norm vector, it is easy to show that $\textbf{n}_k^p$ is still Gaussian distribution whose elements are independently and identically distributed as $\mathcal{CN}(0,\frac{1}{Kp_k^p}{\mathbf{I}}_M)$. The MMSE estimate of channel $\textbf{h}_k$ is given by
{\setlength\abovedisplayskip{3pt}
\setlength\belowdisplayskip{5pt}
\begin{equation}
\hat{\textbf{h}}_k=\frac{{{\alpha _k}Kp_k^p}}{{{\alpha _k}Kp_k^p + {1}}}\textbf{y}_k^p,
\end{equation}}
which follows the distribution of $\mathcal{CN}(\mathbf{0},\sigma_k\mathbf{I})$ with $\sigma _k$  given by
\vspace{-0.05cm}
{\setlength\abovedisplayskip{5pt}
\setlength\belowdisplayskip{-9pt}
\begin{equation}\label{sqdeewferf}
  \sigma _k = \frac{{\alpha _k^2Kp_k^p}}{{{\alpha _k}Kp_k^p + 1}}.
\end{equation}}

According to the property of MMSE estimation, channel estimation error $\tilde{\textbf{h}}_k=\textbf{h}_k-\hat{\textbf{h}}_k$ is independent of $\hat{\textbf{h}}_k$, and follows the  distribution of $\mathcal{CN}(\mathbf{0},\delta_k\mathbf{I}_M)$, where $\delta_k$ is given by
\vspace{-0.05cm}
{\setlength\abovedisplayskip{3pt}
\setlength\belowdisplayskip{-15pt}
\begin{equation}\label{nofewdiwdjn}
\delta_k=\frac{{{\alpha _k}}}{{{\alpha _k}Kp_k^p + 1}}.
\end{equation}}

\subsection{Achievable Data Rate for Massive MIMO URRLC}

By taking into account the number of symbols for pilot transmission, to achieve the decoding error probability of $\varepsilon_k$ for the $k$th device, the instantaneous achievable data rate $R_k$ can be accurately approximated by \cite{mousaei2017optimizing}
\vspace{-0.1cm}
{\setlength\abovedisplayskip{3pt}
\setlength\belowdisplayskip{4pt}
\begin{equation}\label{ratex}
\vspace{-0.05cm}
{R_k} \approx  {\left( {1 - \beta } \right){{\log }_2}(1 + {\gamma _k}) - \sqrt {\frac{{\left( {1 - \beta } \right){V_k}}}{L}} \frac{{{Q^{ - 1}}({\varepsilon _k})}}{{\ln 2}}},
\end{equation}}
\!\!where $\beta$ is equal to $\beta  = {{{K}} \mathord{\left/
 {\vphantom {{{K}} L}} \right.
 \kern-\nulldelimiterspace} L}$, $\gamma_k$ is the signal to interference plus noise ratio (SINR) of the $k$th device, $Q^{-1}$ is the inverse function $Q(x)=\frac{1}{\sqrt{2\pi}}\int_x^\infty e^{-\frac{t^2}{2}} \text{d}t$, and $V_k$ is the channel dispersion given by $V_k=1-(1+\gamma_k)^{-2}$.  As seen from (\ref{ratex}), when the blocklength $L$ approaches infinity, the data rate $R_k$ will approach  $\left( {1 - \beta } \right){\log _2}(1 + \gamma_k )$, which is  the classic Shannon capacity. The second term in (\ref{ratex}) can be interpreted as a penalty on the rate in order to guarantee the decoding error probability $\varepsilon_k$.

In the following, we derive the expression of $\gamma_k$ for two different low-complexity detection schemes: 1) maximum-ratio combining (MRC); 2)  zero-forcing (ZF).

\begin{figure}
\centering
\includegraphics[width=6.5in]{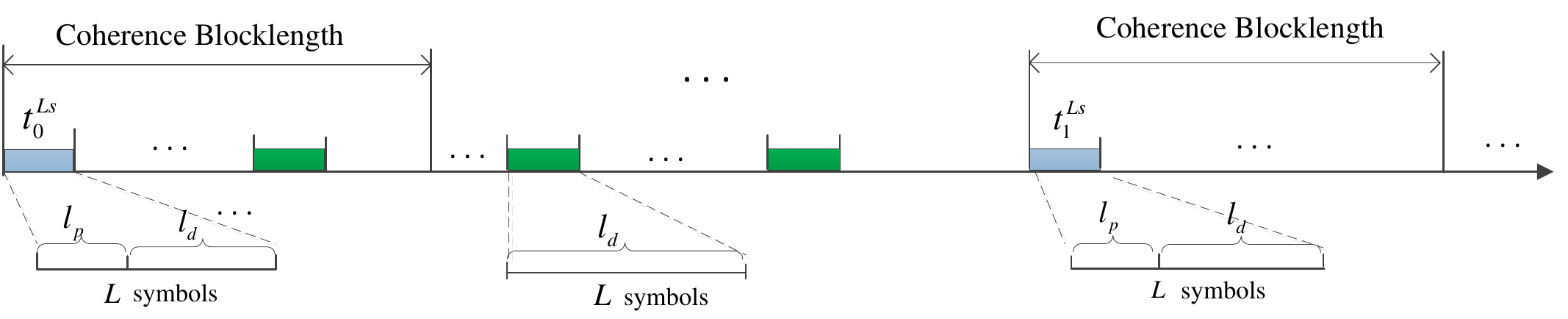}
\vspace{-0.7cm}
\caption{Block diagram for transmission in TDD URLLC Massive MIMO scheme.}
\label{coherencetime}
\vspace{-1cm}
\end{figure}

Define the estimated channels as $\widehat{\textbf{H}}=[\hat{\textbf{h}}_1,\hat{\textbf{h}}_2,\cdots,\hat{\textbf{h}}_K]$ and channel estimation errors as ${\widetilde{\bf H}} = [{{{\tilde{\bf h}}}_1},{{{\tilde{\bf h}}}_2}, \cdots ,{{{\tilde{\bf h}}}_K}]$.
Let $\bf{A}$ be an $M \times K$ linear detection matrix that is based on the estimated channel $\widehat{\textbf{H}}$. By using the linear detection $\bf{A}$, the received signal can be processed as
\vspace{-0.2cm}
\begin{equation}\label{koyloil}
\vspace{-0.3cm}
  {\bf{y}}^D = {{\bf{A}}^H}{\bf{y}}.
\end{equation}
Two conventional low-complexity linear detectors are considered:
\vspace{-0.05cm}
\begin{equation}\label{joihiywgwge}
\vspace{-0.05cm}
 {\bf{A}} = \left\{ \begin{array}{l}
\widehat {\bf{H}},\qquad\qquad\quad  {\rm{for}}\quad {\rm{MRC}}\\
\widehat {\bf{H}}{\left( {{{\widehat {\bf{H}}}^H}\widehat {\bf{H}}} \right)^{ - 1}},\ {\rm{for}}\quad {\rm{ZF}}.
\end{array} \right.
\end{equation}
Then, the processed signal after using the detector is given by
\vspace{-0.1cm}
\begin{eqnarray}
\vspace{-0.35cm}
 {\bf{y}}^D &=& {{\bf{A}}^H}\sum\nolimits_{i \in \mathcal{K}} {{{\bf{h}}_i}\sqrt {p_i^d} } {s_i} + {{\bf{A}}^H}{\bf{n}}.\nonumber\\
   &=& {{\bf{A}}^H}\sum\nolimits_{i \in \mathcal{K}} {{{{\hat{\bf h}}}_i}\sqrt {p_i^d} } {s_i} + {{\bf{A}}^H}\sum\nolimits_{i \in \mathcal{K}} {{{{\tilde{\mathbf h}}}_i}\sqrt {p_i^d} } {s_i} + {{\bf{A}}^H}{\bf{n}},\label{dfwfew}
\vspace{-0.2cm}
\end{eqnarray}
\!\!where the last equality is obtained by using ${{\bf{h}}_k} = {{{\hat{\bf h}}}_k} + {{{\tilde{\bf h}}}_k}$. The detection signal for the $k$th device is given by
\vspace{-0.1cm}
\begin{equation}\label{cxsdnbcblmn}
\vspace{-0.1cm}
  {{\bf{y}}^D_k} = {\mathbf{a}}_k^H{{{\hat{\bf h}}}_k}\sqrt {p_k^d} {s_k} + {\bf{a}}_k^H\sum\nolimits_{i \in \mathcal{K} \setminus k} {{{{\hat{\bf h}}}_i}\sqrt {p_i^d} } {s_i} + {\bf{a}}_k^H\sum\nolimits_{i \in \mathcal{K}} {{{{\tilde{\mathbf h}}}_i}\sqrt {p_i^d} } {s_i} + {\mathbf{a}}_k^H{\mathbf{n}}.
\end{equation}
\!\!where ${\bf{a}}_k$ is the $k$th column of matrix ${\bf{A}}$. Since   $\widehat{\textbf{H}}$ and $\widetilde {\textbf{H}}$ are independent,  ${\bf{a}}_k$ is also independent of $\widetilde {\textbf{H}}$. The CC will treat the estimated channel as the true channel, and the last three terms of (\ref{cxsdnbcblmn}) are regarded as interference and noise. Then, the SINR for the $k$th device $\gamma_k$ is given by
\begin{equation}\label{moinou}
{\gamma _k} = \frac{{p_k^d{{\left| {{\bf{a}}_k^H{{{\hat{\bf h}}}_k}} \right|}^2}}}{{\sum\nolimits_{i \in \mathcal{K}\setminus k} {p_i^d{{\left| {{\bf{a}}_k^H{{{\hat{\bf h}}}_i}} \right|}^2}}  + \sum\nolimits_{i \in \mathcal{K}} {p_i^d{{\left| {{\bf{a}}_k^H{{{\tilde{\bf h}}}_i}} \right|}^2}}  + {{\left\| {{{\bf{a}}_k}} \right\|}^2}}}.
\end{equation}
{\bf{Remark}:} Massive MIMO can offer the channel hardening effect, where the channel variations between different channel fading blocks can be averaged out and the achievable data rate for these fading blocks mainly depend on the large-scale fading, which changes very slowly. As a result, we first derive the lower bound (LB) for the achievable data rate as a function of large-scale channel fading parameters, and then optimize the resource allocation based on the large-scale fading information rather than the small-scale fading, which can significantly \emph{reduce the computational delay} that is beneficial for URLLC applications. In other words, when the large-scale fading parameters of all devices are given, we can use our developed algorithm to find the optimal power allocation, which can be used for consecutive channel fading blocks. The algorithms are needed to be rerun only when the large-scale channel fading parameters have changed, which vary much slowly compared with the small-scale channel fading. To make it more clear, a block diagram scheme is given in Fig.~\ref{coherencetime}, where the large-scale channel gains of any devices vary at $t=t_0^{Ls}$ and $t=t_1^{Ls}$. In general, the channel coherence time is much longer than the packet transmission time as we consider the URLLC services, and the time difference for large-scale fading ($t_1^{Ls}-t_0^{Ls}$) is much larger than the channel coherence time. The power allocation obtained at time $t=t_0^{Ls}$ can be employed for the subsequent transmissions until $t=t_1^{Ls}$, which significantly reduce the computational time. Please note in our scheme, the channel estimation should be performed at the beginning of each channel coherence since the decoding requires the channel state information as shown in (\ref{joihiywgwge}), while the power allocation needs to be updated once the large-scale fading gains change.

Due to the channel hardening effect, in this paper we focus on the ergodic achievable data rate that is defined as ${\bar R_k}={\mathbb{E}}\left\{{R_k}\right\}$, where the expectation is taken over the randomness of $\left\{ {{{{\hat{\bf h}}}_k},{{{\tilde{\bf h}}}_k},\forall k} \right\}$. Unfortunately, the exact average achievable data rate ${\bar R_k}$ with channel uncertainty is not available. In the following, we aim to derive the closed-form expression of the LB of the rate expression, which is more tractable to analyse and optimize. To this end, we first define function $f(x)$  as
{\setlength\abovedisplayskip{3pt}
\setlength\belowdisplayskip{3pt}
\begin{equation}\label{jhjttypij}
f(x) = \ln \left( {1 + \frac{1}{x}} \right) - a\sqrt {\frac{{2x + 1}}{{{{(x + 1)}^2}}}} \ge 0, x> 0
\end{equation}}
\!\!where $a$  is a fixed positive value. In the following, we derive the feasible region of function $f(x)$. Since $f(x) \ge 0$, from (\ref{jhjttypij}) we have
\begin{equation}\label{johtrguhu}
a \le \frac{{(x + 1)\ln \left( {1 + \frac{1}{x}} \right)}}{{\sqrt {2x + 1} }} \buildrel \Delta \over = g(x).
\end{equation}
The first-order derivative of $g(x)$ with respect to $x$ is given by
\begin{equation}\label{dfeagtju}
g'(x) = \frac{{ - 2 - \frac{1}{x} + x\ln \left( {1 + \frac{1}{x}} \right)}}{{{{\left( {2x + 1} \right)}^{\frac{3}{2}}}}} \le \frac{{ - 1 - \frac{1}{x}}}{{{{\left( {2x + 1} \right)}^{\frac{3}{2}}}}} < 0
\end{equation}
\!\!where the first inequality follows by using the relation $\ln \left( {1 + \frac{1}{x}} \right) < \frac{1}{x}$. Hence, $g(x)$ is a monotonically decreasing function of $x$. In addition, $\mathop {\lim }\limits_{x \to 0} g(x) = \infty $ and  $\mathop {\lim }\limits_{x \to \infty } g(x) = 0$, where the latter equation is obtained by using the L'Hospital's rule. Hence, from (\ref{johtrguhu}), we know that the feasible region of $f(x)$ is given by
\vspace{-0.45cm}
\begin{equation}\label{kukgyuk}
\vspace{-0.45cm}
  {\bf\Omega}  = \left\{ {\left. x \right|0 <x \le {g^{ - 1}}(a)} \right\}.
\end{equation}

Then, we have the following lemma.

\emph{\textbf{Lemma 1}}: Function $f(x)$ defined in (\ref{jhjttypij}) is decreasing and convex for $x\in \bf\Omega$.

\emph{\textbf{Proof}}: \upshape Please refer to  Appendix  B in our recent work \cite{hongwcl}. \hfill\rule{2.7mm}{2.7mm}

Based on Lemma 1, we are able to derive the LB of ${\bar R_k}$ in the following. The instantaneous data rate $R_k$ in (\ref{ratex}) can be written as follows:
\begin{equation}\label{kiykgu}
\vspace{-0.05cm}
{R_k} = \frac{{1 - \beta }}{{\ln 2}}{f_k}\left( {\frac{1}{{{\gamma _k}}}} \right),
\end{equation}
where function ${f_k}(\cdot)$ is in the same format as $f(\cdot)$ in (\ref{jhjttypij}), where the parameter $a$ is
\vspace{-0.2cm}
\begin{equation}\label{mknbklnki}
\vspace{-0.3cm}
 {a_k} = {{{Q^{ - 1}}({\varepsilon _k})}}/{\sqrt {L\left( {1 - \beta } \right)} }.
\end{equation}
\!\!In addition, we need to guarantee that $R_k\ge 0$, and thus $\gamma _k\ge {1 \mathord{\left/
 {\vphantom {1 {{g^{ - 1}}({a_k})}}} \right.
 \kern-\nulldelimiterspace} {{g^{ - 1}}({a_k})}}$. By using lemma 1 and Jensen's inequality, we obtain the following LB on the ergodic data rate:
\vspace{-0.05cm}
\begin{equation}\label{igkh}
\vspace{-0.05cm}
{{\bar R}_k} \ge {{\hat R}_k} \buildrel \Delta \over = \frac{{1 - \beta }}{{\ln 2}}{f_k}\left( {{\mathbb{E}}\left\{ {\frac{1}{{{\gamma _k}}}} \right\}} \right).
\end{equation}
In the following theorem, we derive the expression of ${{\hat R}_k}$  for each beamforming solution.

\emph{\textbf{Theorem 1}}:  The ergodic achievable rate for the $k$th device for MRC in finite blocklenghth regime can be lower bounded by:
\vspace{-0.2cm}
\begin{equation}\label{sdeffre}
\vspace{-0.1cm}
  {{\hat R}_k} \buildrel \Delta \over = \frac{{1 - \beta }}{{\ln 2}}{f_k}\left( \frac{1}{{{{\hat \gamma }_k}}}  \right)
\end{equation}
where ${\hat \gamma}_k$ is given by
\begin{equation}\label{jhoyjmkin}
\vspace{-0.1cm}
{{\hat \gamma }_k} = \frac{{p_k^d(M - 1){\sigma _k}}}{{\sum\nolimits_{i \in \mathcal{K} \setminus k} {p_i^d{\sigma _i}}  + \sum\nolimits_{i \in \mathcal{K}} {p_i^d{\delta _i}} + 1}}.
\end{equation}

\emph{\textbf{Proof}}: \upshape Please refer to  Appendix  \ref{proofoftheorem1}. \hfill\rule{2.7mm}{2.7mm}

For the ZF detection, the LB of the ergodic data rate is given by the following theorem.

\emph{\textbf{Theorem 2}}:
The ergodic achievable rate for the $k$th device for ZF in finite blocklength regime is lower bounded by:
\vspace{-0.2cm}
{\setlength\abovedisplayskip{3pt}
\setlength\belowdisplayskip{3pt}
\begin{equation}\label{sdeaedare}
\vspace{-0.3cm}
  {{\hat R}_k} \buildrel \Delta \over = \frac{{1 - \beta }}{{\ln 2}}{f_k}\left( \frac{1}{{{{\hat \gamma }_k}}}  \right)
\end{equation}}
\!\!where ${\hat \gamma }_k$ is given by
\begin{equation}\label{jhdcswefn}
\vspace{-0.2cm}
{{\hat \gamma }_k} = \frac{{(M - K){\sigma _k}p_k^d}}{{\sum\nolimits_{i \in \mathcal{K}} {p_i^d{\delta _i} + 1} }}.
\end{equation}

\emph{\textbf{Proof}}: \upshape Please refer to  Appendix  \ref{proofoftheorem2}. \hfill\rule{2.7mm}{2.7mm}

In the proof of Theorem 1 and Theorem 2, we notice that the above derived LB of the ergodic data rates are valid for any number of antennas ($M>K$ for the ZF). However, the gap between the LB and the actual ergodic data rate is reduced when the number of antennas is  large, which is the case for massive MIMO system. These LBs are commonly used in the literature concerning massive MIMO systems. Hence, we use these lower bounds throughout this paper. In the following, we aim to optimize the pilot power $p_k^p$ and payload power $p_k^d$  to maximize the weighted sum rate. The optimization is performed only when any large-scale fading parameter changes. The simulation results in Section \ref{simulation} also verify the tightness of the derived LBs. Hence, the optimization solutions are applicable for \emph{a large time scale}, which is appealing for URLLC applications.  

\vspace{-0.2cm}
\subsection{Problem Formulation}

In this paper, we jointly optimize the  power allocation for pilot and data transmission of each device for maximizing the weighted sum rate of all devices.  By using Theorem 1, the weighted sum rate maximization problem can be formulated as
\vspace{-0.5cm}
\begin{spacing}{1.1}
\begin{subequations}\label{initial-wse}
\begin{align}
\mathop {\max }\limits_{\{ p_k^p\} ,\{ p_k^d\} } \;\;\;& \sum\nolimits_{k \in \mathcal{K}} {{w_k}{{\hat R}_k}} \\
{\rm{s.t.}}\;\;\;
&  {{\hat R}_k} \ge R_k^{{\rm{req}}},\forall k,\label{ehfgho}\\
&{K}p_k^p + (L - K)p_k^d \le {E_k},\forall k,\label{freocdsdi}
\end{align}
\end{subequations}
\end{spacing}
\noindent where ${{\hat R}_k}$ and ${{\hat \gamma }_k}$ are given in (\ref{sdeffre}) and (\ref{jhoyjmkin}) for MRC and (\ref{sdeaedare}) and (\ref{jhdcswefn}) for ZF, respectively, $w_k$ is the weight of device $k$ used to guarantee the fairness among the devices, constraint (\ref{ehfgho}) denotes the minimum data rate requirement for the $k$-th device, constraint (\ref{freocdsdi}) means the energy constraint for each device.

Power control for weighted sum rate problem with interference is well known to be an NP-hard problem even under perfect CSI \cite{luo2008dynamic}. It becomes more complicated for the more general case with imperfect CSI and finite blocklength. In this paper, we aim for designing efficient algorithms with polynomial-time complexity to solve the weighted sum rate problem with imperfect CSI.

To this end, we first simplify the problem formulation in (\ref{initial-wse}). The first-order derivative of ${{\hat R}_k}$ w.r.t. ${\hat \gamma }_k$ is given by ${{\hat R'}_k} =  - \frac{{\left( {1 - \beta } \right)}}{{\hat \gamma _k^2\ln 2}}{f_k'}\left( {\frac{1}{{{{\hat \gamma }_k}}}} \right) \ge 0$ \footnote{Since ${{\hat R}_k} \ge R_k^{{\rm{req}}}>0$, the feasible ${{\hat \gamma }_k}$  must lie in the range of  ${\bf\Omega}  = \left\{ {\left. {\hat \gamma }_k \right|0 <1/{\hat \gamma }_k \le {g^{ - 1}}(a_k)} \right\}$. Hence, Lemma 1 holds and we have ${f_k'}\left( {\frac{1}{{{{\hat \gamma }_k}}}} \right) \ge 0$.}. Hence, constraint (\ref{ehfgho}) can be  transformed as
\vspace{-0.1cm}
{\setlength\abovedisplayskip{3pt}
\setlength\belowdisplayskip{5pt}
\begin{equation}\label{fhhthtyh}
\vspace{-0.1cm}
 {{\hat \gamma }_k} \ge {1 \mathord{\left/
 {\vphantom {1 {{f_{k}^{ - 1}}\left( {\frac{{R_k^{{\rm{req}}}\ln 2}}{{1 - \beta }}} \right)}}} \right.
 \kern-\nulldelimiterspace} {{f_{k}^{ - 1}}\left( {\frac{{R_k^{{\rm{req}}}\ln 2}}{{1 - \beta }}} \right)}},\forall k.
\end{equation}}
\!To additionally simplify the problem formulation in (\ref{initial-wse}), we introduce auxiliary variables ${\chi _k},\forall k$, and then Problem (\ref{initial-wse}) can be equivalently transformed as follows
\vspace{-0.5cm}
\begin{spacing}{1.0}
\begin{subequations}\label{gsrgtse}
\begin{align}
\mathop {\max }\limits_{\{\chi _k\}, \{ p_k^p\} ,\{ p_k^d\} } \;\;\;& \sum\nolimits_{k \in \mathcal{K}} {{\tilde w}_k}\left[ {\ln (1 + {\chi _k}) - {a_k}G({\chi _k})} \right] \label{jfoweouh}\\
{\rm{s.t.}}\;\;\;& {{\hat \gamma }_k} \ge \chi _k,\forall k,\label{djrdsfrefjgot}\\
&(\ref{fhhthtyh}), (\ref{freocdsdi}),
\end{align}
\end{subequations}
\end{spacing}
\noindent where  ${{\tilde w}_k} = \frac{{\left( {1 - \beta } \right){w_k}}}{{\ln 2}}$ and $G(\chi _k)=\sqrt {1 - \frac{1}{{{{(1 + \chi _k)}^2}}}}$. The equivalence between Problem  (\ref{gsrgtse}) and Problem (\ref{initial-wse}) can be readily proved by using the contradiction method. Problem (\ref{gsrgtse}) and Problem (\ref{initial-wse}) are equivalent in the sense that  they have the same  power allocation solutions and the same objective function (OF) value. Hence, in the following, we focus on solving Problem (\ref{gsrgtse}).

Due to different SINR expressions, in the following two sections, we optimize the power allocation for MRC and ZF, respectively.

\section{Weighted Sum Data Rate for MRC}\label{MRCcase}
\vspace{-0.1cm}
In this section, we aim to deal with weighted sum rate maximization problem for MRC.

\vspace{-0.6cm}
\subsection{Algorithm Design}
\vspace{-0.1cm}
The complicated function $G(\chi _k)$ in  (\ref{jfoweouh})  makes the optimization problem difficult to solve. To resolve this issue, we first study several properties of this function.

\emph{\textbf{Lemma 2}}:
$G({x})$ is a concave function of $x$.

\emph{\textbf{Proof}}: \upshape Please refer to  Appendix  \ref{proofoflemma2}. \hfill\rule{2.7mm}{2.7mm}

For the URLLC applications, the decoding error probability ${\varepsilon _k}$ for each device is much smaller than $0.5$ so that $a_k$ is a strictly positive value, and the OF of Problem (\ref{gsrgtse}) is to maximize the difference of two concave functions. However, due to the non-convex constraints in Problem (\ref{gsrgtse}). This problem does not belong to the class of difference of convex (DC) problem. In addition, due to the additional term of ${a_k}G({\chi _k})$ in the OF of Problem (\ref{gsrgtse}), this problem cannot be solved by dealing with a sequence of geometric programs (GPs) as in \cite{cheng2015uplink}, which considered Shannon capacity formula under the infinite blocklength regime. The intuitive method to solve  Problem (\ref{gsrgtse}) is to approximate function $G(\chi _k)$ as its first-order Taylor series expansion and solve the approximated problem until convergence. However, the approximation is in linear form while the first term in the OF is in log-function form. The successive GP method in \cite{cheng2015uplink} is still not applicable.  To deal with this issue, we approximate function $G(x)$  as a log-function as shown in the following lemma.

\emph{\textbf{Lemma 3}}: For any given  $\tilde x\ge \frac{{\sqrt {17}  - 3}}{4}$,  the following inequality holds:
{\setlength\abovedisplayskip{3pt}
\setlength\belowdisplayskip{3pt}
\begin{equation}\label{frejoig}
  G(x)=\sqrt {1 - \frac{1}{{{{(1 + x)}^2}}}}\le \rho{\rm{ln}}(x)+\eta \buildrel \Delta \over = F(x), \forall x\ge \frac{{\sqrt {17}  - 3}}{4},
\end{equation}}
\!\!where $\rho$ and $\eta$ are given by
{\setlength\abovedisplayskip{3pt}
\setlength\belowdisplayskip{3pt}
\begin{equation}\label{joijojo}
  {\rho} = \frac{{\tilde x}}{{\sqrt {{{\tilde x}^2} + 2\tilde x} }} - \frac{{\tilde x\sqrt {{{\tilde x}^2} + 2\tilde x} }}{{{{\left( {1 + \tilde x} \right)}^2}}},
\end{equation}}
\!\!and
\vspace{-0.1cm}
{\setlength\abovedisplayskip{3pt}
\setlength\belowdisplayskip{3pt}
\begin{equation}\label{swdefefr}
\vspace{-0.2cm}
{\eta} = \sqrt {1 - \frac{1}{{{{\left( {1 + \tilde x} \right)}^2}}}}  - {\rho}\ln (\tilde x).
\end{equation}}
\!\!In addition, we have:
\vspace{-0.1cm}
{\setlength\abovedisplayskip{3pt}
\setlength\belowdisplayskip{3pt}
\begin{equation}
\label{equalA}
\vspace{-0.5cm}
G(\tilde x)=F(\tilde x), G'(\tilde x)=F'(\tilde x),
\end{equation}}
\!\!which means that the approximation $F(x)$ is tight at $x=\tilde x$.

\emph{\textbf{Proof}}: \upshape  Please refer to Appendix  \ref{proofoflemma3}.      \hfill\rule{2.7mm}{2.7mm}

According to (\ref{kukgyuk}), the optimization variable ${\hat \gamma }_k$ should be no smaller than ${1 \mathord{\left/
 {\vphantom {1 {{g^{ - 1}}({a_k})}}} \right.
 \kern-\nulldelimiterspace} {{g^{ - 1}}({a_k})}}$. Note that ${1 \mathord{\left/
 {\vphantom {1 {{g^{ - 1}}({a_k})}}} \right.
 \kern-\nulldelimiterspace} {{g^{ - 1}}({a_k})}}$ is a decreasing function of the decoding error probability and blocklength, while an increasing function of the number of device. Hence, for typical FA cases \cite{yilmaz2015analysis}, where the typical required error probability is lower than ${10^{ - 8}}$, the available channel blocklength is smaller than 200, and the number of devices that should be supported is larger than 5, ${1 \mathord{\left/
 {\vphantom {1 {{g^{ - 1}}({a_k})}}} \right.
 \kern-\nulldelimiterspace} {{g^{ - 1}}({a_k})}}$ is larger than $\frac{{\sqrt {17}  - 3}}{4}$, which implies that Lemma 3 is applicable for our considered optimization problem in (\ref{gsrgtse}).

\begin{figure}
\centering
\includegraphics[width=4.8in]{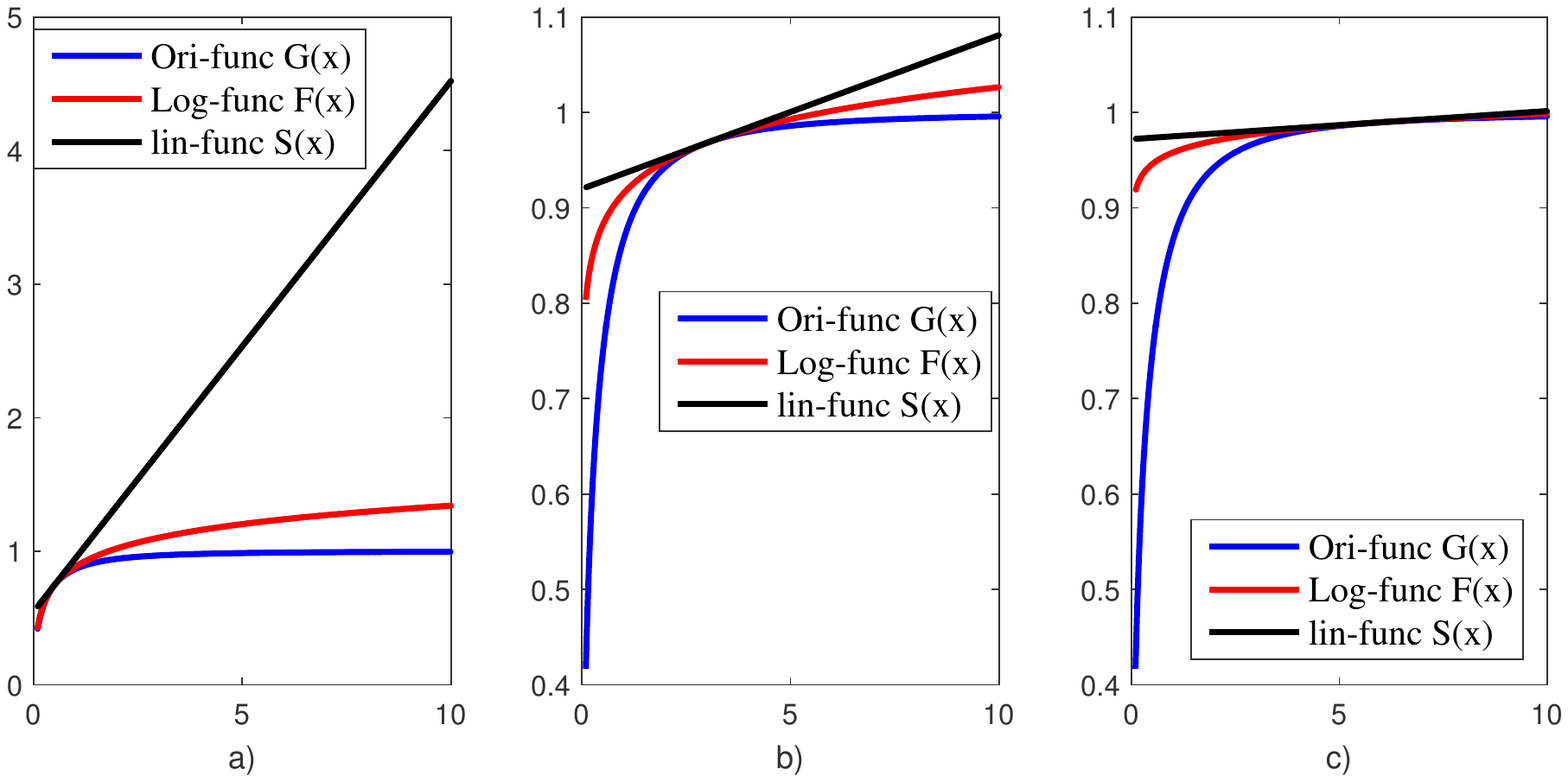}
\vspace{-0.15cm}
\caption{Approximation error for different approximation functions at three different points: a) $\tilde x=0.5$; b) $\tilde x=3$; c) $\tilde x=6$.}
\vspace{-0.1cm}
\label{apprxifun}
\vspace{-0.6cm}
\end{figure}

In Fig.~\ref{Blockdiagram}, we compare the approximation accuracy of function $F(x)$ and the linear approximation function $S(x)$ defined  as follows:
{\setlength\abovedisplayskip{3pt}
\setlength\belowdisplayskip{3pt}
\vspace{-0.1cm}
\begin{equation}\label{xsdgteg}
\vspace{-0.1cm}
  G(x) \le G(\tilde x) + G'(\tilde x)\left( {x - \tilde x} \right) \buildrel \Delta \over = S(x),
\end{equation}}
\!\!where the inequality holds since $G(x)$ is a concave function proved in Lemma 2. It can be observed from Fig.~\ref{apprxifun} that the approximation function $F(x)$ is more accurate than  the linear function $S(x)$ over the whole region of $x$ at different points of $\tilde x$. We can also find that the curve of the log-function $F(x)$ is always above the curve of the linear function $S(x)$, which verifies the correctness of the theoretical conclusion in Lemma 3. The most important advantage of approximating $G(x)$ as the log-function $F(x)$ is that we can transform the original problem into a GP problems that enables us to find the optimal solution.

In the following lemma, we also provide the LB of $\ln (1 + {\chi _k})$, which enables us to develop low-complexity algorithms.

\emph{\textbf{Lemma 4}}: For any given $\tilde x\ge0$,  function $\ln(1+x)$ is lower bounded by
\vspace{-0.2cm}
\begin{equation}\label{adcweafere}
\vspace{-0.3cm}
 \ln (1 + x) \ge {\hat{\rho}}\ln x + \hat{\eta}, \forall x\ge 0,
\end{equation}
where $\hat{\rho}$ and $\hat{\eta}$ are given by
\vspace{-0.15cm}
\begin{equation}\label{nuhohjohj}
\vspace{-0.3cm}
  \hat{\rho} = \frac{{\tilde x}}{{1 + \tilde x}},\hat{\eta} = \ln (1 + \tilde x) - \frac{{\tilde x}}{{1 + \tilde x}}\ln \tilde x.
\end{equation}
In addition, the bound is tight at $x=\tilde x$.

\emph{\textbf{Proof}}: \upshape The proof is similar to those in Lemma 3, and thus omitted for simplicity. \hfill\rule{2.7mm}{2.7mm}

Based on Lemma 3 and Lemma 4, we are now ready to solve Problem (\ref{gsrgtse}). The main idea is to approximate the OF of Problem (\ref{gsrgtse}) as the approximated functions provided in Lemma 3 and Lemma 4, and then solve the approximate problem in an iterative manner. In the following, we provide the details of the iterative algorithm.

First, we denote ${\bf{P}}^{(i)}=\{p_k^{p(i)},p_k^{d(i)},\forall k\}$  as the power allocation in the $i$-th iteration, and the corresponding  $\chi _k$ is given by $\chi _k^{(i)}$. Then, in the $i+1$-th iteration, we can approximate
$G(\chi _k)$ around $\chi _k^{(i)}$ as function $F(\chi _k)= \rho_k^{(i)}{\rm{ln}} \chi_k+\eta_k^{(i)}$, where $\rho_k^{(i)}$ and $\eta_k^{(i)}$ are obtained from (\ref{joijojo}) and (\ref{swdefefr}) respectively with $\tilde x= \chi _k^{(i)}$. In addition, we can approximate  $\ln(1+\chi _k)$ around $\chi _k^{(i)}$ as ${\hat{\rho}_k^{(i)}}\ln \chi _k + {\hat{\eta}_k^{(i)}}$, where ${\hat{\rho}_k^{(i)}}$ and ${\hat{\eta}_k^{(i)}}$ can be obtained from (\ref{nuhohjohj}) with $\tilde x= \chi _k^{(i)}$. By substituting these approximations into  (\ref{jfoweouh}) and recalling that $a_k$ is a positive value, we can obtain the LB of the OF by using Lemma 3 and Lemma 4 as follows
{\setlength\abovedisplayskip{3pt}
\setlength\belowdisplayskip{3pt}
\begin{equation}\label{wedafrgrtj}
  \sum\nolimits_{k \in \mathcal{K}} {{\tilde w}_k}\left[ {\ln (1 + {\chi _k}) - {a_k}G({\chi _k})} \right]  \ge \sum\nolimits_{\mathcal{K}} {{\tilde w}_k}\left[ {{\hat{\rho}_k^{(i)}}\ln \chi _k + {\hat{\eta}_k^{(i)}} - a_k\rho_k^{(i)}{\rm{ln}}\chi_k-a_k\eta_k^{(i)}} \right],
\end{equation}}
\!\!where the equality holds only when $\chi _k=\chi _k^{(i)}$.

Next, we optimize the power allocation to maximize the LB of the OF instead of maximizing (\ref{jfoweouh}) directly. In specific, the LB maximization problem to be solved in the $i+1$-th iteration is formulated as
\vspace{-0.5cm}
\begin{spacing}{1.0}
\begin{subequations}\label{trans-wse}
\begin{align}
\vspace{-0.8cm}
\mathop {\max }\limits_{\{\chi _k\},\{ p_k^p\} ,\{ p_k^d\} } \;\;\;&  \sum\nolimits_{k \in \mathcal{K}} {{\hat w}_k}^{(i)} \ln {\chi _k}\\
{\rm{s.t.}}\;\;\;& (\ref{djrdsfrefjgot}), (\ref{fhhthtyh}), (\ref{freocdsdi}),
\end{align}
\end{subequations}
\end{spacing}
\noindent where ${{\hat w}_k}^{(i)} = {{\tilde w}_k}\hat{\rho}_k^{(i)} - a_k{{\tilde w}_k}\rho_k^{(i)}$ and the constant term in the OF is omitted.

Then, we can transform the above optimization into a GP problem as follows:
\vspace{-0.3cm}
\begin{spacing}{1.2}
\begin{subequations}\label{dwewfrefege}
\vspace{-0.1cm}
\begin{align}
\mathop {\max }\limits_{\{\chi _k\}, \{ p_k^p\} ,\{ p_k^d\} } \;\;\;&  \mathop \prod \nolimits_{k \in \mathcal{K}} \chi _k^{\hat w_k^{(i)}} \label{dsaefaraerg}\\
{\rm{s.t.}}\;\;\;& \sum\limits_{i \in \mathcal{K} \setminus k} \!\! {{\alpha _i}{\alpha _k}{K}{\chi _k}p_k^pp_i^d}  \!\!+\!\! \sum\limits_{i \in \mathcal{K}} {{\alpha _i}{\chi _k}p_i^d}  \!+\! {\chi _k}{\alpha _k}{K}p_k^p \!+\! {\chi _k} \!\le\! (M - 1){K}\alpha _k^2p_k^pp_k^d,\forall k, \label{djrdwWAFDGTrefjgot}\\
&{\chi _k} \ge {1 \mathord{\left/
 {\vphantom {1 {{f_{k}^{ - 1}}\left( {\frac{{R_k^{{\rm{req}}}\ln 2}}{{1 - \beta }}} \right)}}} \right.
 \kern-\nulldelimiterspace} {{f_{k}^{ - 1}}\left( {\frac{{R_k^{{\rm{req}}}\ln 2}}{{1 - \beta }}} \right)}}, \forall k, (\ref{freocdsdi}). \label{djrdsqwdefjgot}
\end{align}
\end{subequations}
\end{spacing}
\noindent Although GP is not a convex optimization problem, it can be equivalently transformed into a convex optimization problem by applying a logarithmic change of variables. Therefore, the globally optimal solution of Problem (\ref{dwewfrefege}) can be obtained by using the interior-point method \cite{boyd2004convex}. Some software packages that can solve GP problem are MOSEK package and  CVX \cite{grant2014cvx}.

Based on the above discussion, the iterative algorithm to solve Problem (\ref{gsrgtse}) is given in  Algorithm \ref{algorithmiterSCA}.

\vspace{-0.3cm}
\begin{algorithm}
\caption{Iterative algorithm for solving Problem (\ref{gsrgtse}) for MRC }\label{algorithmiterSCA}
\begin{algorithmic}[1]
\STATE Initialize  iteration number $i=1$, error tolerance $\xi$. Initialize a feasible power allocation $ \{{p_k^{p}}^{(0)},{p_k^{d}}^{(0)},\forall k\}$, calculate  $\{\chi _k^{(0)}, \rho_1^{(0)},  \rho_2^{(0)},  {{\hat w}_k}^{(0)}, \forall k\}$, and calculate the OF of Problem (\ref{gsrgtse}), denoted as ${\rm{Obj}}^{(0)}$.
 \STATE With given $\{\chi _k^{(i-1)}, \rho_1^{(i-1)}, \rho_2^{(i-1)}, {{\hat w}_k}^{(i-1)}, \forall k\}$,  solve Problem (\ref{dwewfrefege}) by using the CVX package to obtain $ \{p_k^{p(i)},p_k^{d(i)}, \chi _k^{(i)}, \forall k\}$.
 \STATE Update $\{ \rho_k^{(i)}, \rho_k^{(i)}, {{\hat w}_k}^{(i)}, \forall k\}$;
 \STATE Calculate the new OF ${\rm{Obj}}^{(i)}$. If  ${{\left| {{\rm{Ob}}{{\rm{j}}^{(i)}} - {\rm{Ob}}{{\rm{j}}^{(i - 1)}}} \right|} \mathord{\left/
 {\vphantom {{\left| {{\rm{Ob}}{{\rm{j}}^{(i)}} - {\rm{Ob}}{{\rm{j}}^{(i - 1)}}} \right|} {{\rm{Ob}}{{\rm{j}}^{(i)}}}}} \right.
 \kern-\nulldelimiterspace} {{\rm{Ob}}{{\rm{j}}^{(i)}}}} < \xi $, terminate.  Otherwise, set $i \leftarrow i + 1$, go to step 2.
\end{algorithmic}
\end{algorithm}

\vspace{-0.6cm}
\subsection{Algorithm Analysis}
\vspace{-0.1cm}

\subsubsection{Initialization of Algorithm \ref{algorithmiterSCA}}
As shown in Step 1 of Algorithm \ref{algorithmiterSCA}, one has to find a feasible initial power allocation in order to make the algorithm work. Note that randomly selecting a set of power allocation solutions that satisfy the per-device energy constraints may not satisfy their minimum SINR requirements. Hence, one has to carefully choose the initial power allocation. In the following, we provide one alternative method to find the initial power allocation solutions.

Inspired by the user selection problem formulation in \cite{matskani2008convex,pan2017joint}, we construct the following alternative optimization problem by introducing an   auxiliary variable $\varphi$:
\vspace{-0.6cm}
\begin{spacing}{1.2}
\begin{subequations}\label{initdhwifuhise}
\begin{align}
\mathop {\max }\limits_{\varphi, \{ p_k^p\} ,\{ p_k^d\} } \;\;\;& \varphi \\
{\rm{s.t.}}\;\;\;& {{\hat \gamma }_k} \ge { {\varphi \mathord{\left/
 {\vphantom {1 {{f_{k}^{ - 1}}\left( {\frac{{R_k^{{\rm{req}}}\ln 2}}{{1 - \beta }}} \right)}}} \right.
 \kern-\nulldelimiterspace} {{f_{k}^{ - 1}}\left( {\frac{{R_k^{{\rm{req}}}\ln 2}}{{1 - \beta }}} \right)}}},\forall k, (\ref{freocdsdi}).\label{djcdeejgot}
\end{align}
\end{subequations}
\end{spacing}
\noindent Obviously, Problem (\ref{initdhwifuhise}) is always feasible since at least $\{\varphi=0,  p_k^p=0 , p_k^d=0, \forall k \}$ is a feasible solution. It can be readily verified that the original Problem (\ref{initial-wse}) is feasible if the optimal $\varphi\ge 1$, and the output power allocation can be adopted as the initial input for Algorithm \ref{algorithmiterSCA}. In this paper, we assume that Problem (\ref{initial-wse}) is always feasible and the optimal $\varphi$ in Problem (\ref{initdhwifuhise}) is always no smaller than one. Problem (\ref{initdhwifuhise}) can also be transformed into a GP problem, where the globally optimal solution can be obtained. The details are omitted here due to the limited space.

\subsubsection{Convergence Analysis}

In this part, we analyze the convergence of Algorithm \ref{algorithmiterSCA}. In the following, we show that
${\rm{Obj}}^{(i)}\le {\rm{Obj}}^{(i+1)}$.

Since $\{\chi _k^{(i+1)}, \forall k\}$ is the optimal solution of Problem  (\ref{trans-wse}) in the $i+1$-th iteration, we have
\begin{equation}\label{dewf}
\begin{array}{l}
\quad \sum\nolimits_{k \in \mathcal{K}} {{{\tilde w}_k}} \left[ {\hat\rho _k^{(i)}\ln \left( {\chi _k^{(i + 1)}} \right) + \hat\eta _k^{(i)} - {a_k}\rho _k^{(i)}{\rm{ln}}(\chi _k^{(i + 1)}) - {a_k}\eta _k^{(i)}} \right] \\
\ge \sum\nolimits_{k \in \mathcal{K}} {{{\tilde w}_k}} \left[ {\hat\rho _k^{(i)}\ln \left( {\chi _k^{(i)}} \right) + \hat\eta _k^{(i)} - {a_k}\rho _k^{(i)}{\rm{ln}}(\chi _k^{(i)}) - {a_k}\eta _k^{(i)}} \right]= {\rm{Obj}}^{(i)}.
\end{array}
\end{equation}
By using inequality (\ref{wedafrgrtj}) with $\chi _k=\chi _k^{(i+1)}$, we have
\begin{equation}\label{sefwjutej}
\begin{array}{l}
\quad\sum\nolimits_{k \in \mathcal{K}}  {{\tilde w}_k}\left[ {\ln (1 + \chi _k^{(i + 1)}) - {a_k}G(\chi _k^{(i + 1)})} \right]\\
 \ge \sum\nolimits_{k \in \mathcal{K}}  {{\tilde w}_k}\left[ {\hat\rho_k^{(i)}\ln (\chi _k^{(i + 1)}) + \hat\eta_k^{(i)} - {a_k}\rho_k^{(i)}{\rm{ln}}(\chi _k^{(i + 1)}) - {a_k}\eta_k^{(i)}} \right].
\end{array}
\end{equation}
By combining (\ref{dewf}) and (\ref{sefwjutej}), we have
\begin{equation}\label{johohuoehw}
{\rm{Ob}}{{\rm{j}}^{(i + 1)}} \ge \sum\nolimits_{k \in \mathcal{K}} {{{\tilde w}_k}} \left[ {\hat\rho_k^{(i)}\ln (\chi _k^{(i + 1)}) + \hat\eta_k^{(i)} - {a_k}\rho_k^{(i)}{\rm{ln}}(\chi _k^{(i + 1)}) - {a_k}\eta_k^{(i)}} \right] \ge {\rm{Ob}}{{\rm{j}}^{(i)}}.
\end{equation}

In addition, since each device has its own energy constraint, the OF value of Problem (\ref{initial-wse}) has upper bound.  As a result, Algorithm \ref{algorithmiterSCA} is guaranteed to converge.

\subsubsection{Solution Analysis}

Since the original problem (\ref{initial-wse}) is non-convex, the globally optimal solution cannot be obtained in general. However, by using the similar proof as in Appendix B in \cite{pan2017joint}, we can prove that Algorithm \ref{algorithmiterSCA} can converge to the Karush-Kuhn-Tucker (KKT) point of Problem (\ref{initial-wse}).   The converged solution only depends on the initial input of Algorithm \ref{algorithmiterSCA}. Simulation results show that the algorithm almost converges to the same solution with different initial input solutions.

\subsubsection{Complexity Analysis}

The main complexity mainly lies in solving a GP problem in each iteration. In \cite{boyd2007tutorial}, the authors claimed that the GP problem can be efficiently solved by using the standard interior point methods with a worst-case polynomial-time complexity.
The upper bound of the total number of Newton steps in the interior point method does not depend on the number of variables, or the number of constraints. The derived upper bound shows that the barrier method converges linearly. By carefully choosing the parameters, the bound on the number of Newton steps can grow as $\sqrt{m}$ instead of $m$, where $m$ is the number of  constraints \cite{shi2008rate}.  In addition, simulation results show that Algorithm 1 converges rapidly, which means that Algorithm \ref{algorithmiterSCA} can converge to a local optimal solution with a polynomial time complexity.

\section{Weighted Sum Data Rate for ZF}\label{ZFcase}
In this section, we aim to deal with weighted sum rate maximization problem for ZF. Due to different expressions of the SINR of MRC and ZF, some derivations in MRC cannot be directly applied in the ZF case. In the following, we develop an efficient algorithm to solve Problem (\ref{gsrgtse}) for the ZF case.
\vspace{-0.4cm}
\subsection{Algorithm Design}
\vspace{-0.1cm}
By substituting (\ref{sqdeewferf}) and (\ref{nofewdiwdjn}) into (\ref{jhdcswefn}), the SINR in the ZF case can be reformulated as
\begin{equation}\label{uoeaughgdf}
 {{\hat \gamma }_k} = \frac{{(M - K)\alpha _k^2Kp_k^pp_k^d\prod\nolimits_{i \in \mathcal{K}} {(1 + {\alpha _i}Kp_i^p)} }}{{\left( {1 + {\alpha _k}Kp_k^p} \right)\left( {\sum\nolimits_{i\in \mathcal{K}} {p_i^d{\alpha _i}\prod\nolimits_{j \ne i} {(1 + {\alpha _j}Kp_j^p)}  + \prod\nolimits_{i \in \mathcal{K}} {(1 + {\alpha _i}Kp_i^p)} } } \right)}},\forall k.
\end{equation}
Unfortunately, since the nominator of ${{\hat \gamma }_k}$ in (\ref{uoeaughgdf}) is a posynomial function, the SINR constraint in (\ref{djrdsfrefjgot}) cannot be transformed into the format as that in (\ref{djrdwWAFDGTrefjgot}), in which the right hand side is a monomial function. Hence, the problem cannot be directly transformed into a GP problem. To solve this issue, we introduce Theorem 3 as follows.

\emph{\textbf{Theorem 3}}: For any given vector ${ \breve{\bf x}}=\{\breve x_1,\cdots,\breve x_K\}$ with $\breve x_i\ge 0,\forall n$, function $W({\bf{x}}) = \prod\nolimits_{i \in \mathcal{K}} {(1 + {x_i})} $ is lower bounded by
\vspace{-0.1cm}
{\setlength\abovedisplayskip{3pt}
\setlength\belowdisplayskip{3pt}
\begin{equation}\label{jjorghot}
\vspace{-0.1cm}
W({\bf{x}}) = \prod\nolimits_{i\in\mathcal{K}} {(1 + {x_i})}  \ge \lambda \prod\nolimits_{i\in \mathcal{K}} {x_i^{{\tau _i}}}  \buildrel \Delta \over = Y({\bf{x}})
\end{equation}}
\!\!where $\lambda$ and $\tau _i,\forall i$ are given by
{\setlength\abovedisplayskip{3pt}
\setlength\belowdisplayskip{4pt}
\begin{equation}\label{hihuiuewui}
  \lambda  = \frac{{\prod\nolimits_{i\in\mathcal{K}} {(1 + {{\breve x}_i})} }}{\prod\nolimits_{i\in \mathcal{K}} {\breve x_i^{{\tau _i}}} },{\tau _i} = \frac{{{{\breve x}_i}}}{{1 + {{\breve x}_i}}},\forall i,
\end{equation}}
\!\!and ${\bf{x}}$ is given by ${\bf{x}}=\{x_1,\cdots,x_K\}$.

In addition, we have:
\vspace{-0.25cm}
{\setlength\abovedisplayskip{-3pt}
\setlength\belowdisplayskip{-3pt}
\begin{equation}\label{jitiijyiij}
\vspace{-0.45cm}
W({\breve{\bf x}}) = Y({\breve{\bf x}}),\nabla W({\breve{\bf x}}) = \nabla Y({\breve{\bf x}}).
\end{equation}}
\!\!where $\nabla W(\bf{x})$ and $\nabla Y(\bf{x})$ denote the gradient of function $W(\cdot)$ and $Y(\cdot)$ w.r.t. $\bf{x}$, respectively.

\emph{\textbf{Proof}}: \upshape Please refer to Appendix  \ref{proofoftheorem3}. \hfill\rule{2.7mm}{2.7mm}

Based on Theorem 3, we replace the polynomial functions in the nominator of ${{\hat \gamma }_k}$ in (\ref{uoeaughgdf}) with their best local monomial approximations by employing Theorem 3 \footnote{The best local monomial approximations means that the approximation function should satisfy three conditions as specified in  Section IV-A of \cite{chiang2005power}.}. In specific, we denote ${\bf{P}}^{(n)}=\{p_k^{p(n)},p_k^{d(n)},\forall k\}$  as the power allocation in the $n$-th iteration, and the corresponding  $\chi _k$ is given by $\chi _k^{(n)}$. Then, in the $n+1$-th iteration, we approximate the term ${\prod\nolimits_{i \in \mathcal{K}} {(1 + {\alpha _i}Kp_i^p)} }$ in the nominator of (\ref{uoeaughgdf}) by its best local monomial approximations, which is given by function $Y({\bf{x}})$ in Theorem 3 with $x_i={\alpha _i}Kp_i^p$:
\vspace{-0.15cm}
{\setlength\abovedisplayskip{3pt}
\setlength\belowdisplayskip{3pt}
\begin{equation}\label{dewfrf}
\prod\nolimits_{i\in \mathcal{K}} {(1 + {\alpha _i}Kp_i^p)}  \ge {\lambda ^{(n)}}\prod\nolimits_{i \in \mathcal{K}} {{{\left( {{\alpha _i}Kp_i^p} \right)}^{{\tau _i^{(n)}}}}},
\end{equation}}
\!\!where ${\lambda ^{(n)}}$ and ${\tau _i^{(n)}}, \forall i$ are given in (\ref{hihuiuewui}) with ${\tilde x}_i={\alpha _i}Kp_i^{p(n)}, \forall i$. Then, we focus on the following constraint instead of the original SINR constraint in (\ref{djrdsfrefjgot}):
\vspace{-0.2cm}
\begin{align}\label{ilhiuytrew}
\vspace{-0.2cm}
&{\chi _k}\left( {1 + {\alpha _k}Kp_k^p} \right)\left( {\sum\nolimits_{i \in \mathcal{K}} {p_i^d{\alpha _i}\prod\nolimits_{j \ne i} {(1 + {\alpha _j}Kp_j^p)}  + \prod\nolimits_{i\in \mathcal{K}} {(1 + {\alpha _i}Kp_i^p)} } } \right)\nonumber\\
 \le &(M - K)\alpha _k^2Kp_k^pp_k^d{\lambda ^{(n)}}\prod\nolimits_{i \in \mathcal{K}} {{{\left( {{\alpha _i}Kp_i^p} \right)}^{{\tau _i}^{(n)}}}}.
\end{align}
Note that the left hand side (LHS) of (\ref{ilhiuytrew}) is a posynomial function, and the right hand side (RHS) becomes a monomial function. In addition, the LHS is no larger than RHS. Hence, constraint (\ref{ilhiuytrew}) satisfies the conditions for a problem to be a GP problem \cite{boyd2004convex}.

Then, by using the same method as in the MRC case to deal with the OF, in the $n+1$-th iteration, we aim to solve the following GP problem:
\vspace{-0.1cm}
\begin{subequations}\label{dwewsge}
\begin{align}
\vspace{-0.2cm}
\mathop {\max }\limits_{\{\chi _k\}, \{ p_k^p\} ,\{ p_k^d\} } \;\;\;&  \mathop \prod \nolimits_{k \in \mathcal{K}} \chi _k^{\hat w_k^{(n)}} \label{dsaraerg}\\
{\rm{s.t.}}\;\;\;&      (\ref{freocdsdi}), (\ref{ilhiuytrew}) \label{djrdTrefjgot}\\
&{{\hat \gamma }_k} \ge  {1 \mathord{\left/
 {\vphantom {1 {{f_{k}^{ - 1}}\left( {\frac{{R_k^{{\rm{req}}}\ln 2}}{{1 - \beta }}} \right)}}} \right.
 \kern-\nulldelimiterspace} {{f_{k}^{ - 1}}\left( {\frac{{R_k^{{\rm{req}}}\ln 2}}{{1 - \beta }}} \right)}}, \forall k, \label{djrdseefjgot}
\end{align}
\end{subequations}
where the parameters ${\hat w_k^{(n)}}$'s are the same as those in the MRC case. This problem can be efficiently solved by using CVX \cite{grant2014cvx}.

Based on the above discussion, the iterative algorithm to solve Problem (\ref{gsrgtse}) for the case of ZF is provided in  Algorithm \ref{algorithmiterzf}.

\vspace{-0.2cm}
\begin{algorithm}
\caption{Iterative algorithm for solving Problem (\ref{gsrgtse}) for ZF }\label{algorithmiterzf}
\begin{algorithmic}[1]
\STATE Initialize  iteration number $n=1$, error tolerance $\xi$. Initialize a feasible power allocation $ \{p_k^{p(0)},p_k^{d(0)},\forall k\}$, calculate  $\{\chi _k^{(0)}, \rho_1^{(0)},  \rho_2^{(0)},  {{\hat w}_k}^{(0)}, \lambda ^{(0)}, \tau_k^{(0)}, \forall k\}$, and calculate the OF of Problem (\ref{gsrgtse}), denoted as ${\rm{Obj}}^{(0)}$.
 \STATE With given $\{\chi _k^{(n-1)}, \rho_1^{(n-1)}, \rho_2^{(n-1)}, {{\hat w}_k}^{(n-1)}, \lambda ^{(n-1)}, \tau_k^{(n-1)} \forall k\}$,  solve Problem (\ref{dwewsge}) by using the CVX package to obtain $ \{p_k^{p(n)},p_k^{d(n)}, \chi _k^{(n)}, \forall k\}$.
 \STATE Update $\{ \rho_1^{(n)}, \rho_2^{(n)}, {{\hat w}_k}^{(n)}, \lambda ^{(n)}, \tau_k^{(n)}, \forall k\}$;
 \STATE Calculate new OF ${\rm{Obj}}^{(n)}$. If  ${{\left| {{\rm{Ob}}{{\rm{j}}^{(n)}} - {\rm{Ob}}{{\rm{j}}^{(n - 1)}}} \right|} \mathord{\left/
 {\vphantom {{\left| {{\rm{Ob}}{{\rm{j}}^{(n)}} - {\rm{Ob}}{{\rm{j}}^{(n - 1)}}} \right|} {{\rm{Ob}}{{\rm{j}}^{(n)}}}}} \right.
 \kern-\nulldelimiterspace} {{\rm{Ob}}{{\rm{j}}^{(n)}}}} < \xi $, terminate.  Otherwise, set $n \leftarrow n + 1$, go to step 2.
\end{algorithmic}
\end{algorithm}

\vspace{-0.3cm}
\subsection{Algorithm Analysis}
\vspace{-0.1cm}
Algorithm \ref{algorithmiterzf} can be analyzed similar to Algorithm \ref{algorithmiterSCA} except the convergence analysis since the problems to be solved in each iteration of Algorithm \ref{algorithmiterzf} do not have the same set of constraints.

In the following, we prove that the solution obtained in the $n$-th iteration is also feasible for the problem to be solved in the $n+1$-th iteration. We only need to check  constraint (\ref{ilhiuytrew})  since the other two constraints are the same in each iteration.

Let us denote $\{\chi _k^{(n)},p_k^{p(n)},p_k^{d(n)},\forall k\}$ as the optimal solution in the $n$-th iteration. Then, it is also a feasible solution, and we have
{\setlength\abovedisplayskip{3pt}
\setlength\belowdisplayskip{3pt}
\begin{align}\label{ilhytrew}
&\chi _k^{(n)}\left( {1 + {\alpha _k}Kp_k^{p(n)}} \right)\left( {\sum\nolimits_{i\in \mathcal{K}} {p_i^{d(n)}{\alpha _i}\prod\nolimits_{j \ne i} {(1 + {\alpha _j}Kp_j^{p(n)})}  + \prod\nolimits_{i \in \mathcal{K}} {(1 + {\alpha _i}Kp_i^{p(n)})} } } \right)\nonumber\\
 \le &(M - K)\alpha _k^2Kp_k^{p(n)}p_k^{d(n)}{\lambda ^{(n - 1)}}\prod\nolimits_{i \in \mathcal{K}} {{{\left( {{\alpha _i}Kp_i^{p(n)}} \right)}^{\tau _i^{(n - 1)}}}}.
\end{align}}
\indent By using (\ref{dewfrf}), we have
{\setlength\abovedisplayskip{3pt}
\setlength\belowdisplayskip{3pt}
\vspace{-0.1cm}
\begin{equation}\label{ehgurghuo}
 \prod\nolimits_{i \in \mathcal{K}} {(1 + {\alpha _i}Kp_i^{p(n)})}  \ge {\lambda ^{(n - 1)}}\prod\nolimits_{i \in \mathcal{K}} {{{\left( {{\alpha _i}Kp_i^{p(n)}} \right)}^{\tau _i^{(n - 1)}}}}.
\end{equation}}
\!\!In addition, by using (\ref{jitiijyiij}) in Theorem 3 and (\ref{ehgurghuo}), we have
{\setlength\abovedisplayskip{3pt}
\setlength\belowdisplayskip{3pt}
\begin{equation}\label{ewfegt}
{\lambda ^{(n)}}\prod\nolimits_{i \in \mathcal{K}} {{{\left( {{\alpha _i}Kp_i^{p(n)}} \right)}^{\tau _i^{(n )}}}}=\prod\nolimits_{i \in \mathcal{K}} {(1 + {\alpha _i}Kp_i^{p(n)})} \ge {\lambda ^{(n - 1)}}\prod\nolimits_{i \in \mathcal{K}} {{{\left( {{\alpha _i}Kp_i^{p(n)}} \right)}^{\tau _i^{(n - 1)}}}}.
\end{equation}}
\!\!Finally, by combining (\ref{ilhiuytrew}) and (\ref{ewfegt}), we have
{\setlength\abovedisplayskip{3pt}
\setlength\belowdisplayskip{3pt}
\begin{align}\label{kojiphjihtj}
&\chi _k^{(n)}\left( {1 + {\alpha _k}Kp_k^{p(n)}} \right)\left( {\sum\nolimits_{i \in \mathcal{K}} {p_i^{d(n)}{\alpha _i}\prod\nolimits_{j \ne i} {(1 + {\alpha _j}Kp_j^{p(n)})}  + \prod\nolimits_{i \in \mathcal{K}} {(1 + {\alpha _i}Kp_i^{p(n)})} } } \right)\nonumber\\
 \le &(M - K)\alpha _k^2Kp_k^{p(n)}p_k^{d(n)}{\lambda ^{(n)}}\prod\nolimits_{i \in \mathcal{K}} {{{\left( {{\alpha _i}Kp_i^{p(n)}} \right)}^{\tau _i^{(n)}}}}.
\end{align}}
\!\!Hence, $\{\chi _k^{(n)},p_k^{p(n)},p_k^{d(n)},\forall k\}$ is also a feasible solution in the $n+1$-th iteration.  Then, by using the similar proof as in the case of MRC, we can also prove that Algorithm \ref{algorithmiterzf} is guaranteed to converge. By using a similar method, we also prove that this algorithm will converge to a feasible solution of Problem (\ref{gsrgtse}).


\section{Simulation Results}\label{simulation}
In this section, we provide simulation results to demonstrate the effectiveness of our proposed algorithms for industrial automation systems. The channel path loss is modeled as
$PL=35.3+37.6{\rm{log}}_{10}d$ (dB) \cite{access2010further}, and the small-scale fading is modeled as Rayleigh fading with zero mean and unit variance. Unless otherwise specified, the simulation parameters are set as follows: number of transmit antennas of $M=100$, number of devices of $K=10$, channel bandwidth of $B=0.2$ MHz, noise power spectral density of -174 dBm/Hz, decoding error probability of ${\varepsilon _k}=10^{-9}, \forall k$, number of transmit antennas of $M=100$, and maximum transmission duration of $0.5$ ms. The maximum blocklength is calculated as $L=BT=100$. The other parameters are specified in each figure. The energy constraint for each device is assumed to be equal, i.e., $E_k=E, \forall k$, and each device has the same data rate targets, $R_k^{\rm{req}}=R^{\rm{req}}, \forall k$. The weights for each device are uniformly generated within $[0,1]$.
\vspace{-0.3cm}
\subsection{Tightness of the Date Rate LB}

\begin{figure}[h]
\begin{minipage}[t]{0.49\linewidth}
\centering
\includegraphics[width=3.2in]{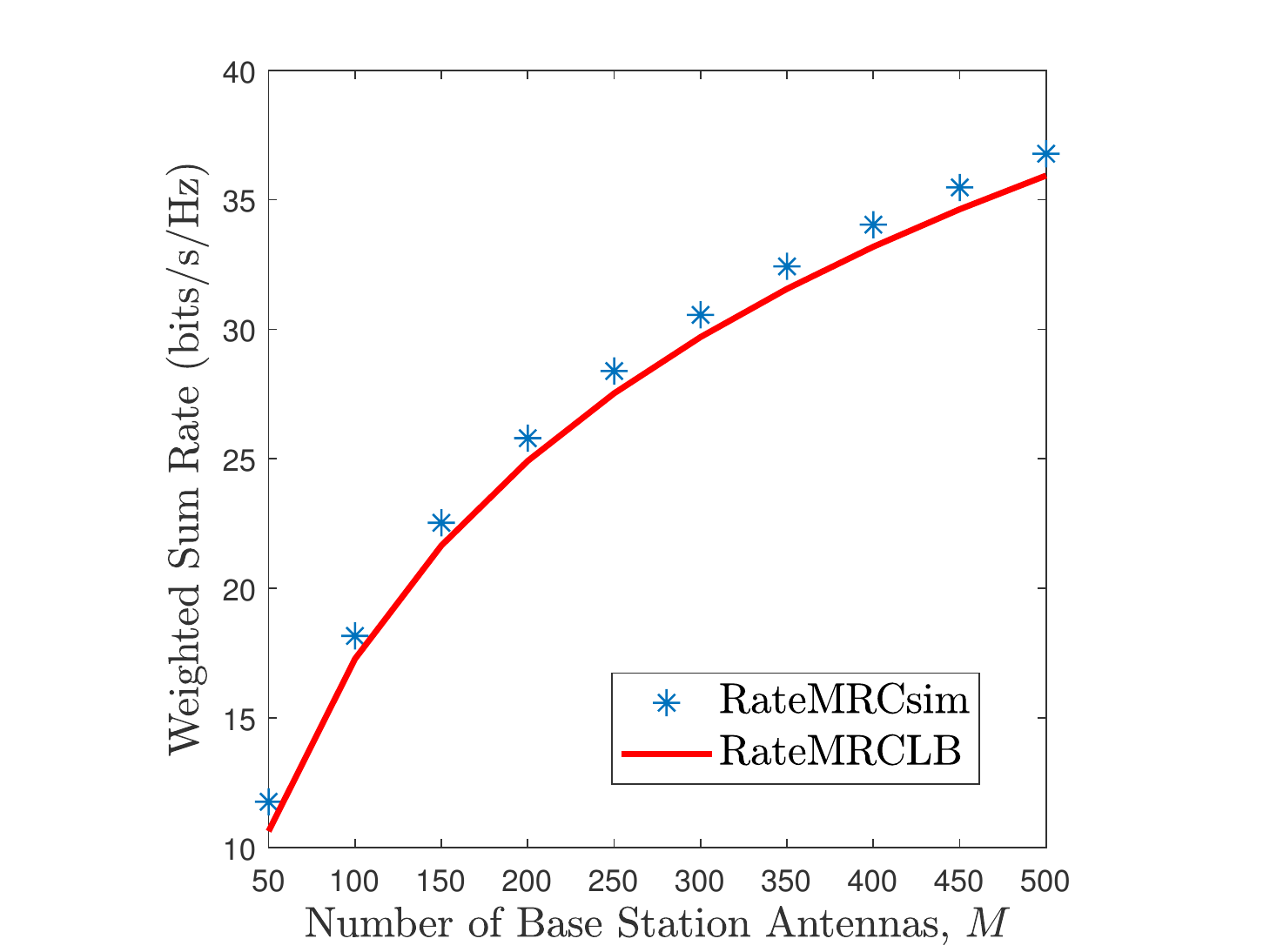}
\vspace{-0.2cm}
\caption{Tightness of the derived data rate LB for the MRC case. }
\label{fig3}\vspace{-0.7cm}
\end{minipage}%
\hfill
\begin{minipage}[t]{0.49\linewidth}
\centering
\includegraphics[width=3.2in]{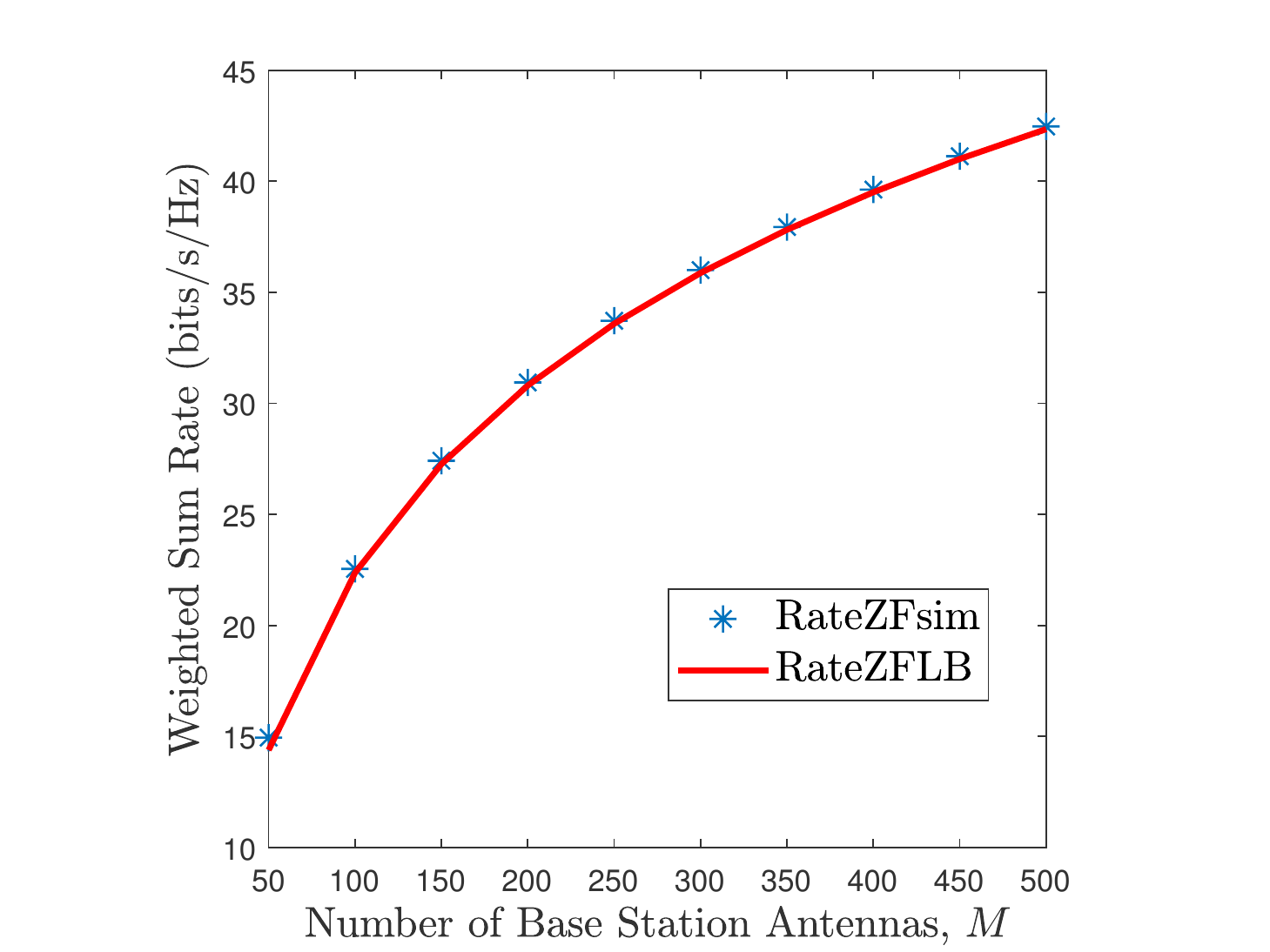}
\vspace{-0.2cm}
\caption{Tightness of the derived data rate LB for the ZF case.}
\label{fig4}\vspace{-0.7cm}
\end{minipage}%
\hfill
\end{figure}

In Fig.~\ref{fig3} and Fig.~\ref{fig4}, we investigates the tightness of the LB derived for the cases of MRC and ZF, respectively. The simulation results are obtained through the Monte-Carlo simulation by averaging over 5000 random channel generations. It is observed that the data rate LB is tight for both cases of MRC and ZF for any number of transmit antennas. Interestingly, the curves for the ZF case are almost overlapped with each other. This verifies that the data rate LBs derived in  Theorem 1 and Theorem 2  are suitable for optimization instead of directly optimizing the complicated expectation expression.

\vspace{-0.6cm}

\subsection{Convergence Behaviour of the Proposed Algorithms}

\begin{figure}[h]
\begin{minipage}[t]{0.49\linewidth}
\centering
\includegraphics[width=3.2in]{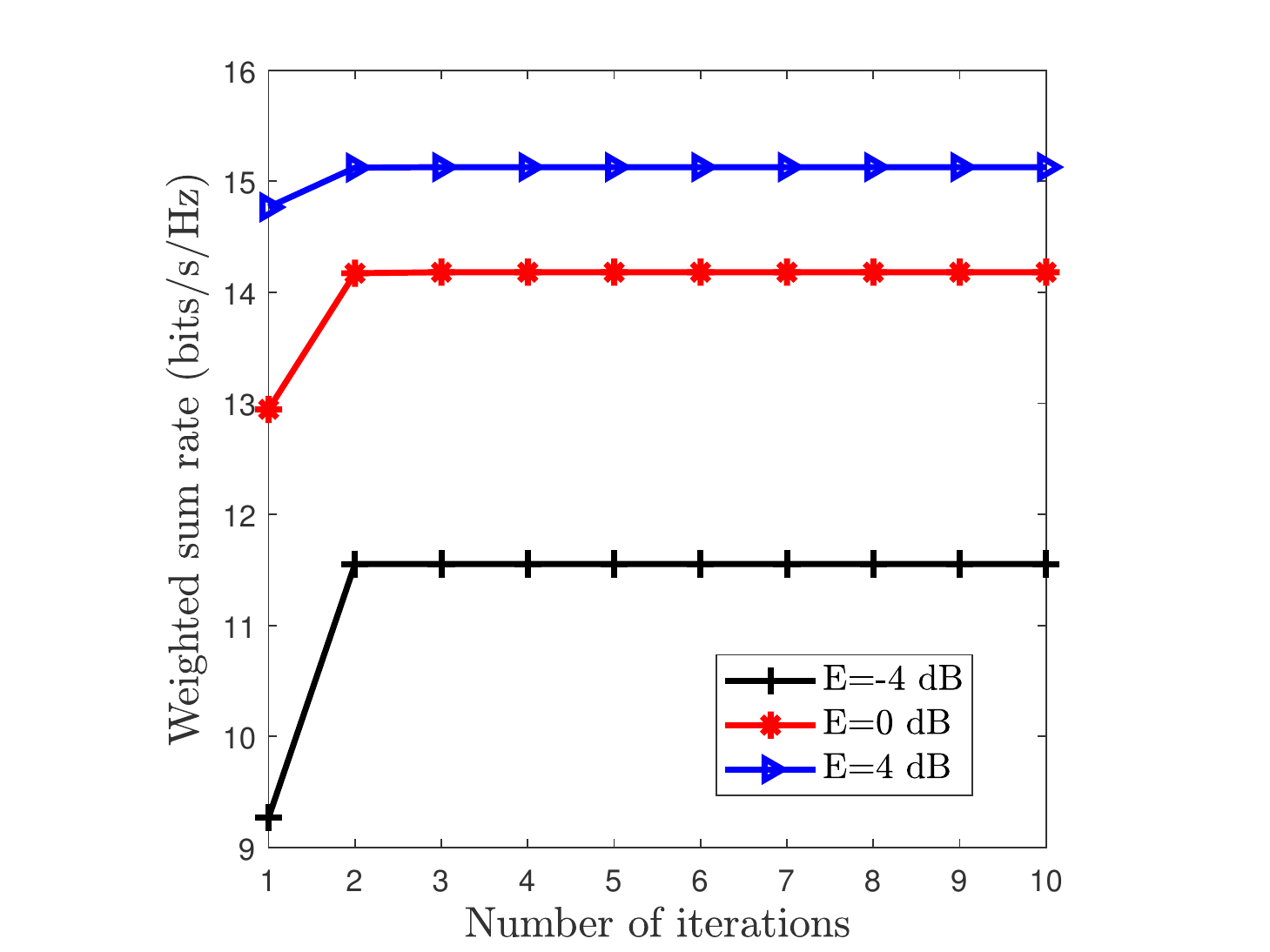}
\vspace{-0.2cm}
\caption{Convergence behaviour of the proposed algorithm for the MRC case. }
\label{fig5}\vspace{-0.8cm}
\end{minipage}%
\hfill
\begin{minipage}[t]{0.49\linewidth}
\centering
\includegraphics[width=3.2in]{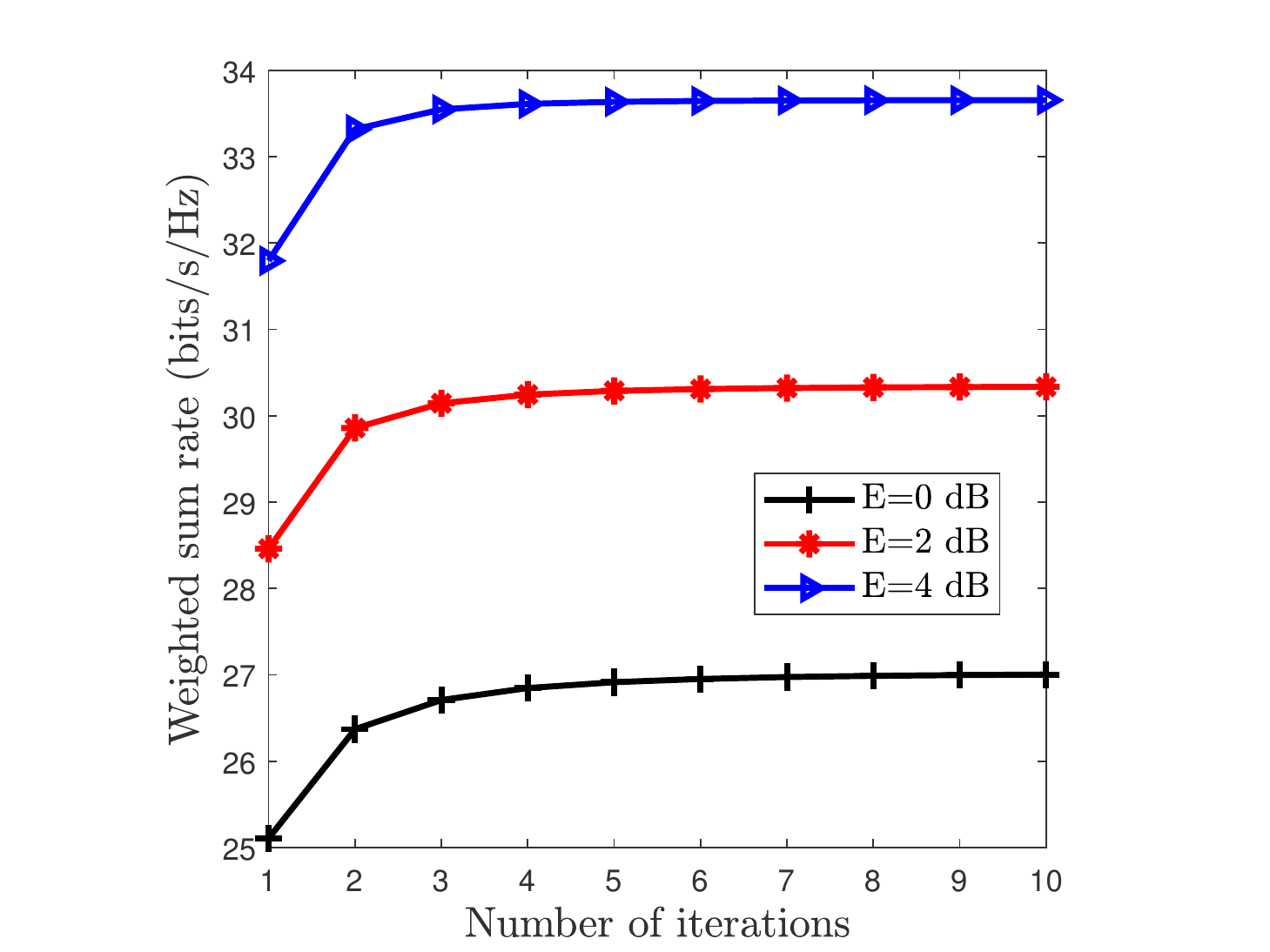}
\vspace{-0.2cm}
\caption{Convergence behaviour of the proposed algorithm for the ZF case.}
\label{fig6}\vspace{-0.8cm}
\end{minipage}%
\hfill
\end{figure}

Fig.~\ref{fig5} and Fig.~\ref{fig6} investigate  the impact of the energy limit at each device on the performance of the proposed Algorithm 1 for the MRC case and Algorithm 2 for the ZF case, respectively. From these two figures, we can observe that both algorithms converge rapidly for various energy limits and only 2 or 3 iterations are sufficient for the algorithms to converge. This demonstrates the low complexity of the proposed algorithms.

\vspace{-0.6cm}

\vspace{0.2cm}
\subsection{Performance Comparison}
In this subsection, we compare the proposed algorithms with the following algorithms:
\begin{itemize}
  \item \textbf{Upper bound:} In this method, Shannon capacity is adopted for optimization in Problem (\ref{gsrgtse}). In other words, the penalty term $a_kG({\chi _k})$  is set to zero in both the OF and rate constraint (\ref{ehfgho}), i.e., $a_kG({\chi _k})=0, \forall k$. This method provides the upper bound of the average weighted sum rate performance in the IIoT networks.
  \item \textbf{Conventional alg.}: As in \cite{Ghanemarix}, the solution obtained from the upper bound is applied in (\ref{sdeffre}) and  (\ref{sdeaedare}) for the cases of MRC and ZF respectively by considering the penalty term $a_kG({\chi _k})$. That means that the upper bound is used for obtaining solutions, but the achievable data rate under finite blocklength is used for performance evaluation.
  \item \textbf{Fixed pilot power alg.}: In this scheme, we only optimize the payload power $p_k^{d}$ while fixing the pilot power as $p_k^{p}=E/L$. This algorithm is provided to show the benefits of jointly optimizing the
      pilot power and payload power.
\end{itemize}
The following results are obtained by averaging over 100 Monte-Carlo simulations where in each snapshot the devices are randomly generated in the cell. For each snapshot, if the device's achievable data rate cannot achieve its rate targets, we set the corresponding data rate to zero. The proposed algorithm is denoted as `Proposed alg.' in the following figures.

In Fig.~\ref{fig7} and Fig.~\ref{fig8}, we show the average weighted sum rate versus the energy limit at each device, $E$. The rate targets for MRC and ZF are set as $R^{\rm{req}}=1$ bit/s/Hz and $R^{\rm{req}}=4$ bit/s/Hz, respectively.  As seen from these figures, the system performance increases with the available energy at each device since the SINR at each device is increased. Note that the weighted sum rate achieved by some algorithms may approach zero, which means the power allocation solution is infeasible. As expected, the upper bound has the best performance since the penalty terms are not considered. Furthermore, the `conventional alg.' based on Shannon capacity formula has higher probability to violate the data rate requirement, especially for samll $E$. Therefore, Shannon capacity cannot be employed for the transmission design of URLLC for industrial applications, in particular when the energy limit is small as the QoS target cannot be guaranteed. However, for the large value of $E$, the SINR value for each device is very high and then the penalty term $a_kG({\chi _k})$ can be ignored when comparing with the first term of ${\ln}(1+{\chi _k})$. Hence, the `conventional alg.' will achieve  similar performance as that of the proposed algorithm. As shown in Fig.~\ref{fig7} and Fig.~\ref{fig8}, by jointly optimizing the  payload power and pilot power, the proposed algorithm is superior over the `Fixed pilot power alg.', which only optimizes the  payload power. The performance gain is obvious when the energy limit is low, especially for the case of ZF. This means the need of jointly optimizing pilot and  payload power at low energy limit. This can be explained as follows. When the energy limit is low, the system performance is limited by channel estimation procedure. Through the joint power allocation, some power can be borrowed from that for data transmission to enhance the channel estimation accuracy, and thus increases the weighted sum rate. However, in the high energy $E$, the accuracy of channel estimation is already enough, and additional joint power control bring marginal performance gain. Another interesting observation is that the performance gain of the proposed algorithm over `Fixed pilot power alg.' is much smaller for the MRC case  than the ZF case.  This may be due to the fact that the CSI is more important when using it for removing multiple device interference in the ZF case.

\begin{figure}
\begin{minipage}[t]{0.49\linewidth}
\centering
\includegraphics[width=3.2in]{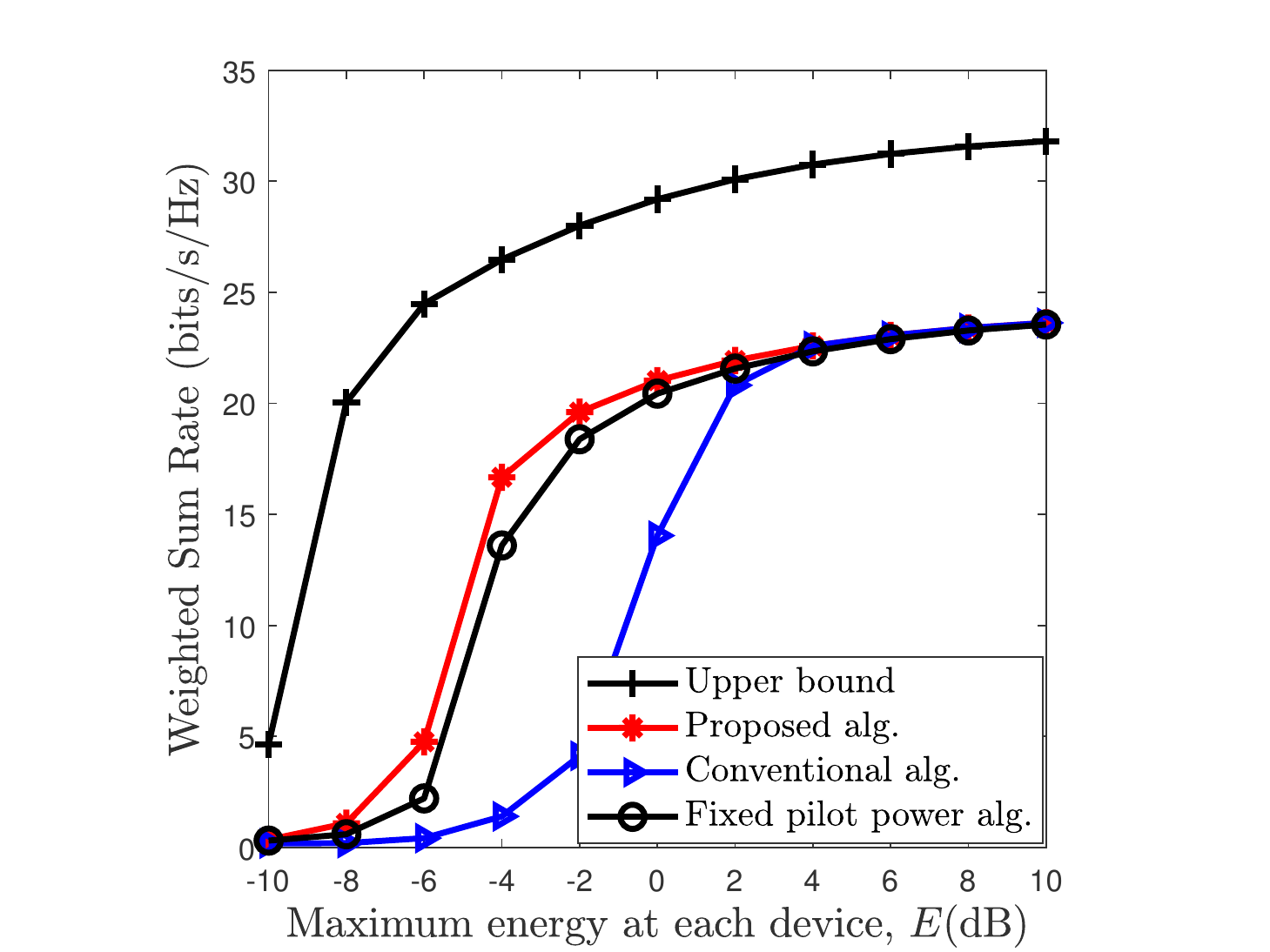}
\vspace{-0.2cm}
\caption{Average weighted sum rate vs. maximum energy limit $10{\rm{log}}_{10}E$ for various schemes for the case of MRC. }
\label{fig7}\vspace{-0.7cm}
\end{minipage}%
\hfill
\begin{minipage}[t]{0.49\linewidth}
\centering
\includegraphics[width=3.2in]{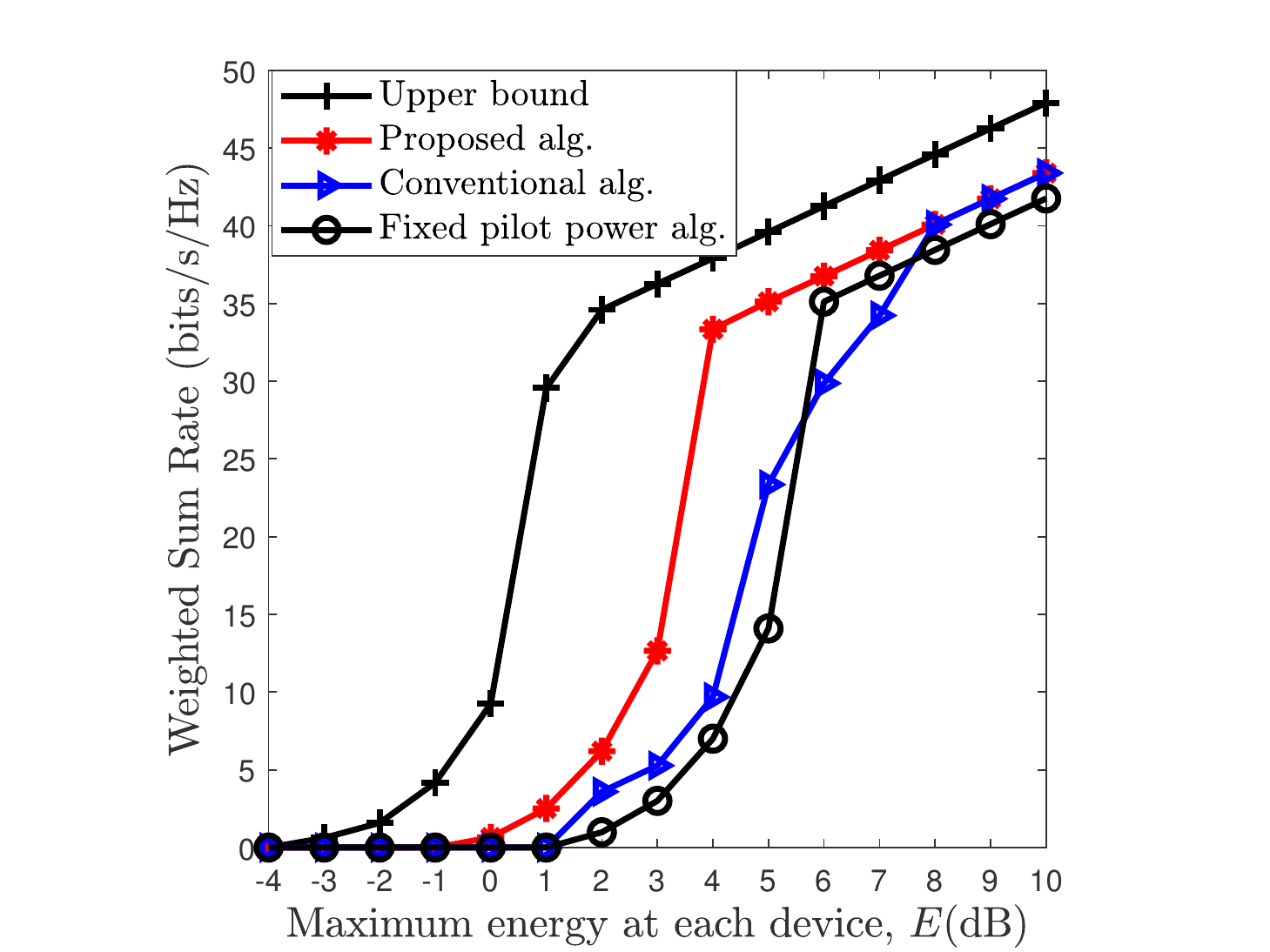}
\vspace{-0.2cm}
\caption{Average weighted sum rate vs. maximum energy limit $10{\rm{log}}_{10}E$ for various schemes for the case of ZF.}
\label{fig8}\vspace{-0.7cm}
\end{minipage}%
\hfill
\end{figure}

\begin{figure}
\begin{minipage}[t]{0.49\linewidth}
\centering
\includegraphics[width=3.2in]{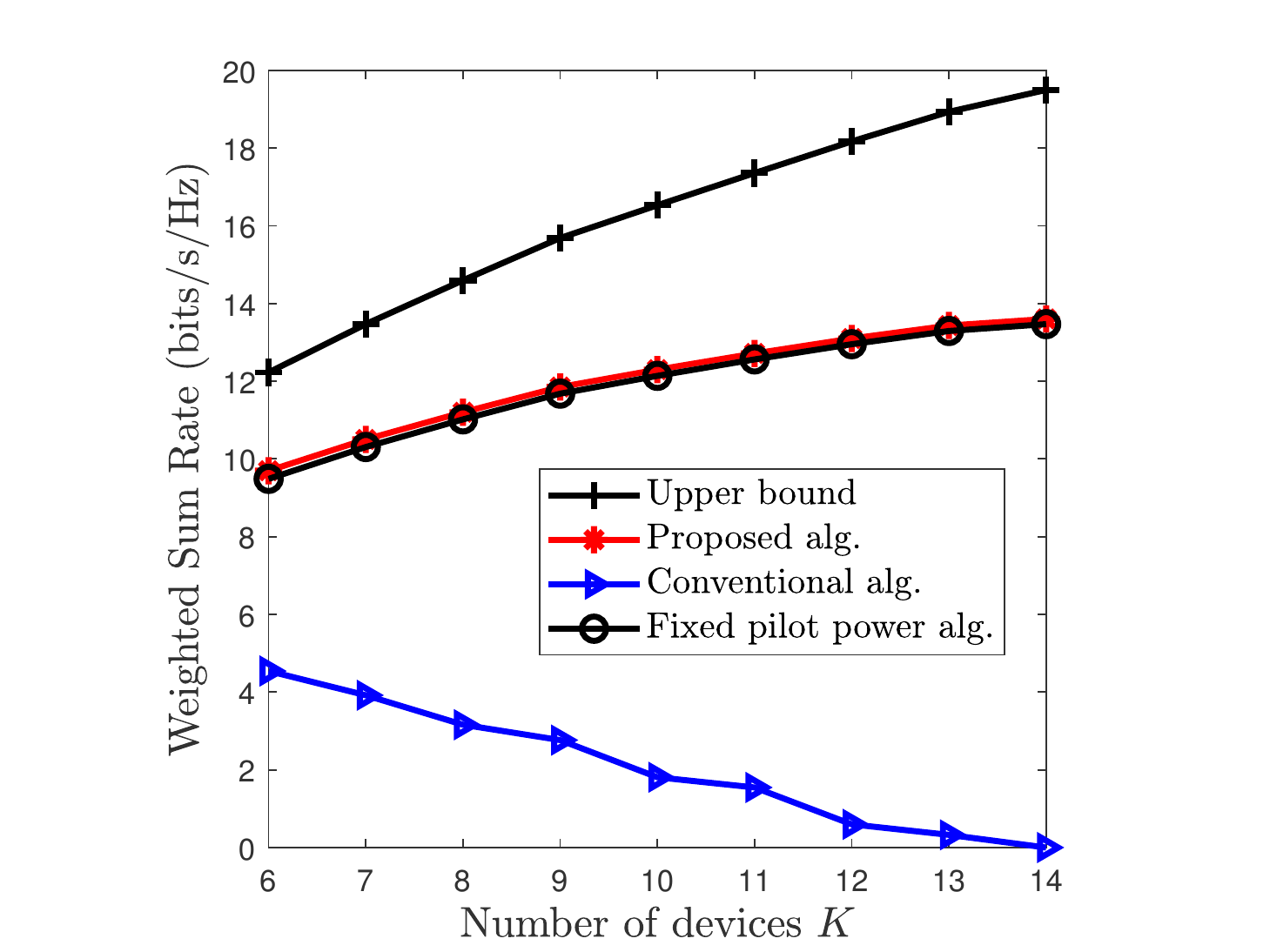}
\vspace{-0.2cm}
\caption{Average weighted sum rate vs. the number of devices for various schemes for the case of MRC. }
\label{fig9}\vspace{-0.7cm}
\end{minipage}%
\hfill
\begin{minipage}[t]{0.49\linewidth}
\centering
\includegraphics[width=3.2in]{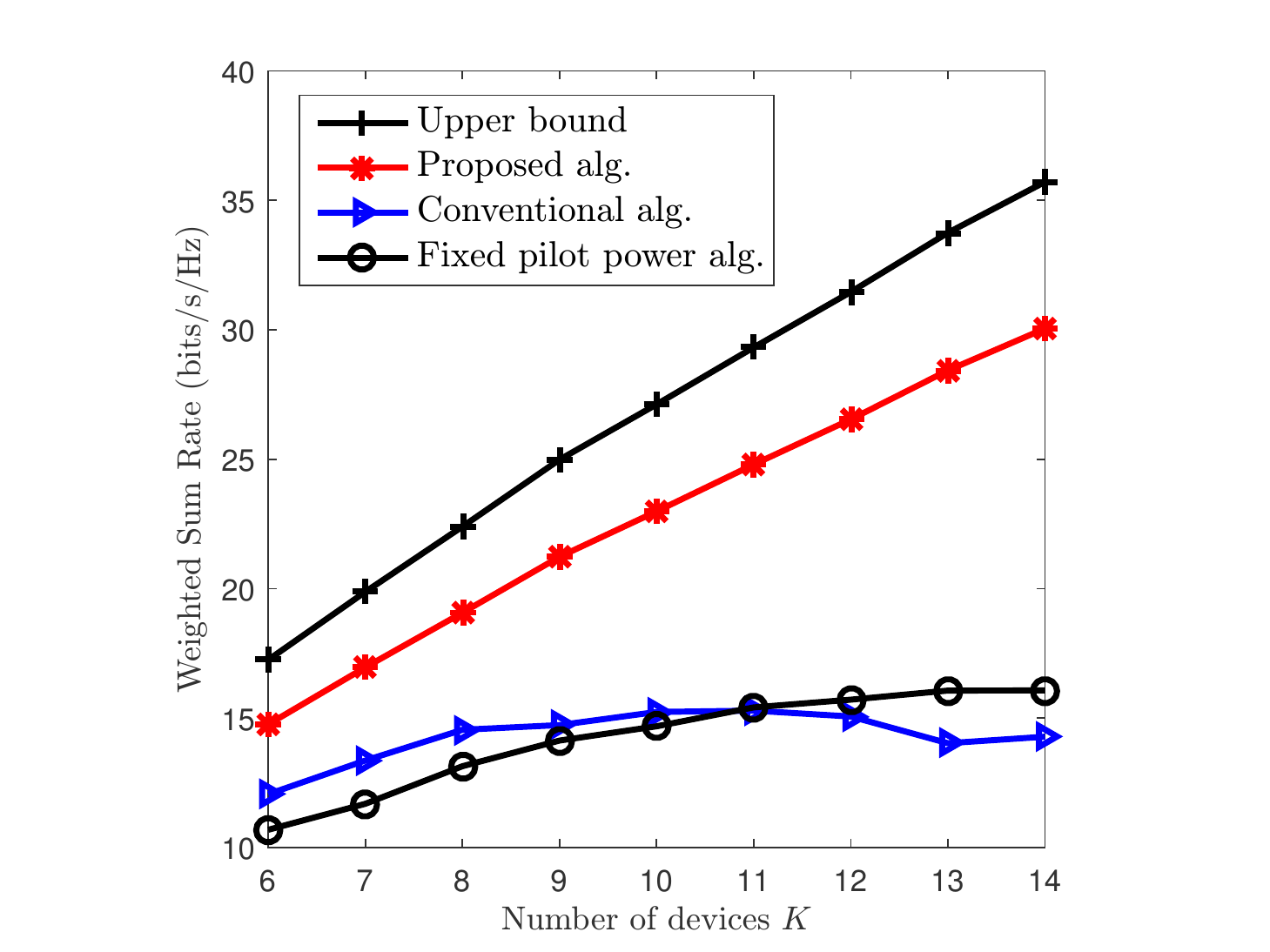}
\vspace{-0.2cm}
\caption{Average weighted sum rate vs. the number of devices  for various schemes for the case of ZF.}
\label{fig10}\vspace{-0.7cm}
\end{minipage}%
\hfill
\end{figure}

Fig.~\ref{fig9} and Fig.~\ref{fig10} show the average weighted sum rate versus the number of devices for various schemes. The rate targets for MRC and ZF are set as $R^{\rm{req}}=1$ bit/s/Hz and $R^{\rm{req}}=2$ bit/s/Hz, respectively. The energy limit for  MRC and ZF are set as $E=2$ and $E=1$, respectively. It is observed from Fig.~\ref{fig9} and Fig.~\ref{fig10} that the average weighted sum rate achieved by all the algorithms except the `Conventional alg.' increases with the number of devices since these schemes fully exploit the multi-device diversity. For the MRC case, the performance of the `Conventional alg.' decreases with the number of devices. The main reason is that the design is based on Shannon capacity formula, which does not take into consideration the effect of short blocklength on the achievable data rate. As a result, the probability that the achieved data rate violates the rate requirement will increase with the number of devices. It is interesting to observe that `Fixed pilot power alg.' has similar performance as the proposed algorithm for the case of MF. However, for the case of ZF, the proposed algorithm significantly outperforms `Fixed pilot power alg.', and the performance gain is increasing with the number of devices. This again reveals the importance of joint power optimization in the case of ZF. For the ZF case, the weighted sum rate achieved by `Conventional alg.'   first increases with the number of devices and then decreases with it.

\begin{figure}
\begin{minipage}[t]{0.49\linewidth}
\centering
\includegraphics[width=3.2in]{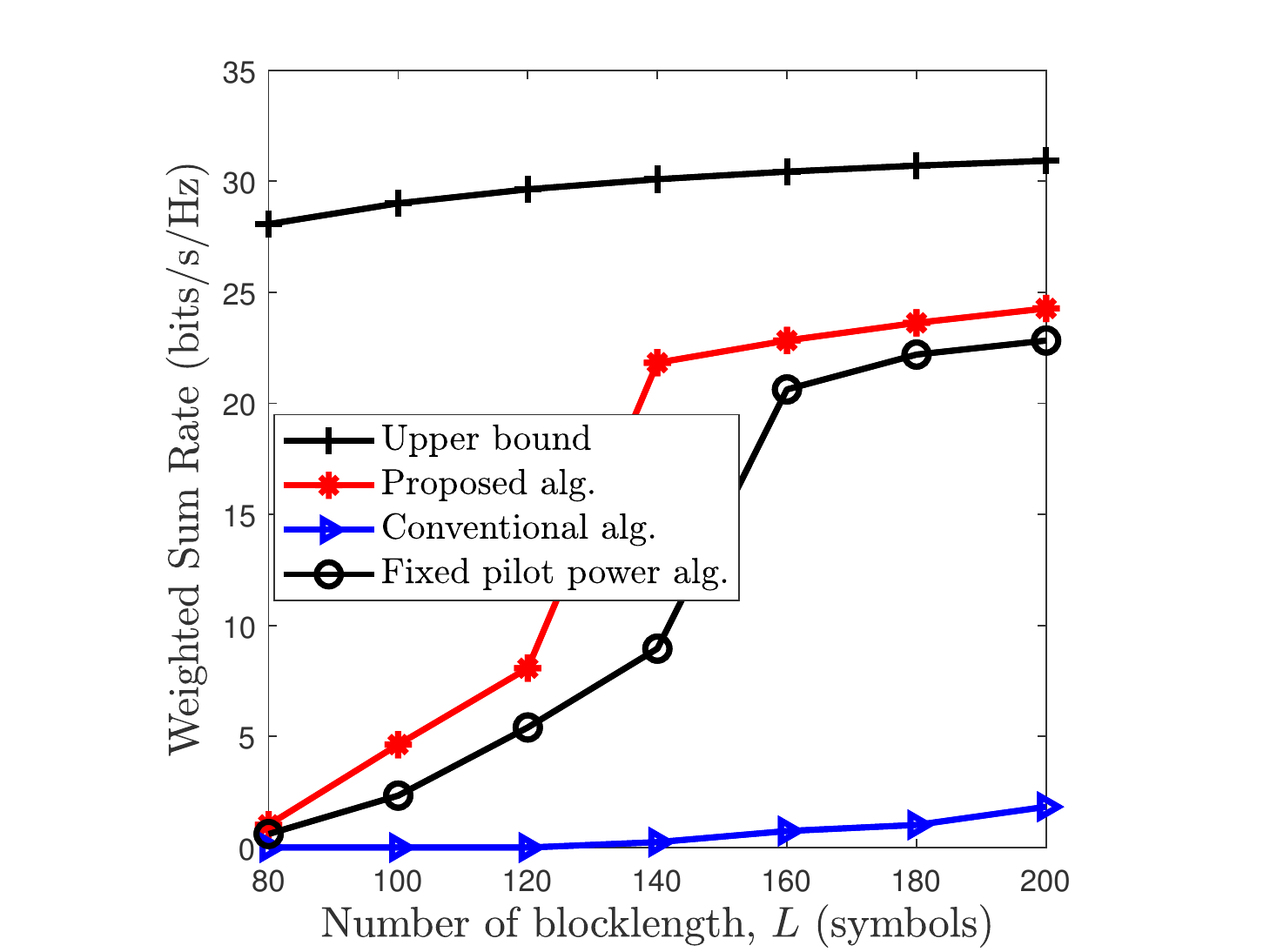}
\vspace{-0.2cm}
\caption{Average weighted sum rate vs. blocklength  for various schemes for the case of MRC. }
\label{fig11}\vspace{-1cm}
\end{minipage}%
\hfill
\begin{minipage}[t]{0.49\linewidth}
\centering
\includegraphics[width=3.2in]{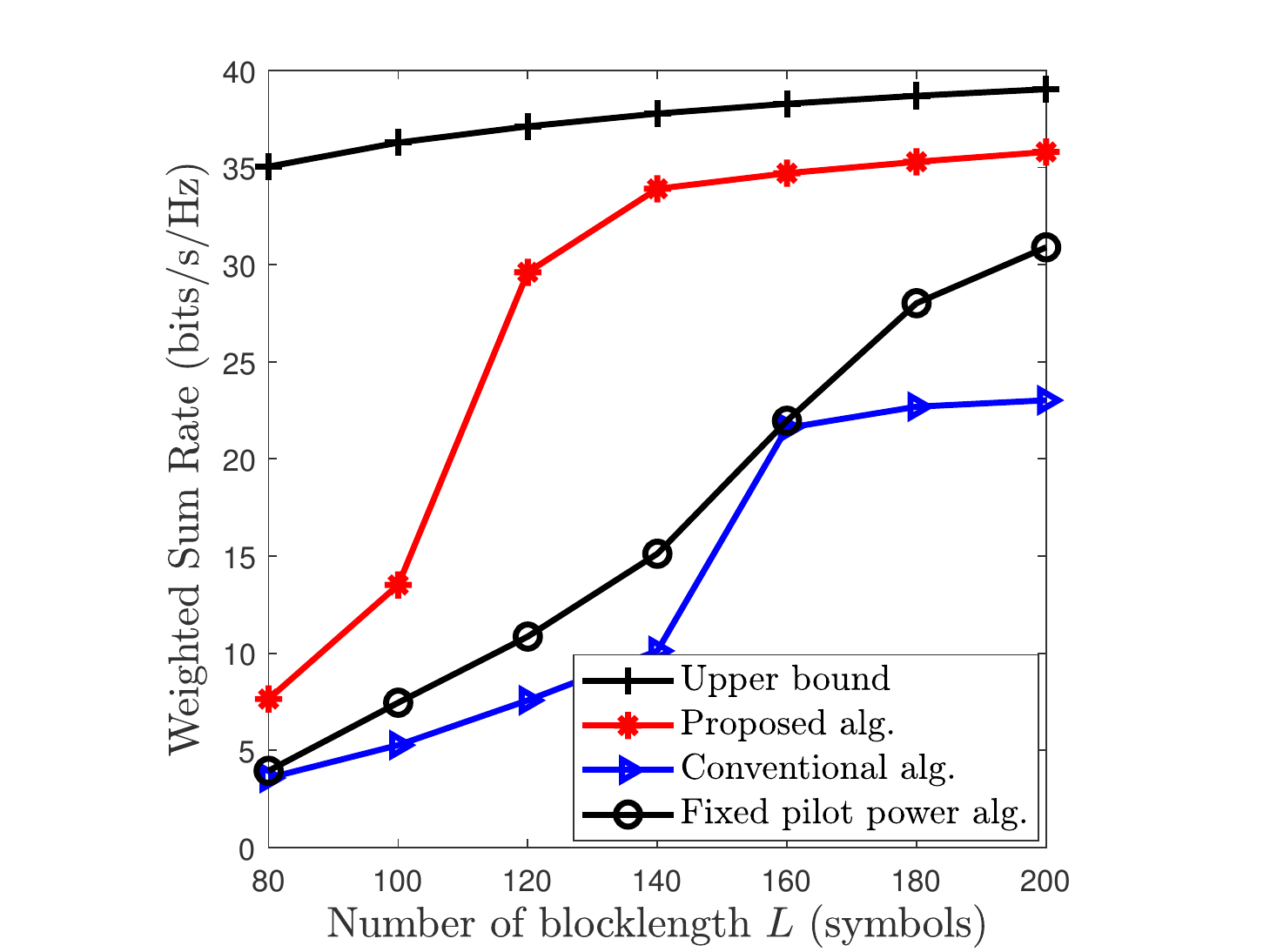}
\vspace{-0.2cm}
\caption{Average weighted sum rate vs. blocklength  for various schemes for the case of ZF.}
\label{fig12}\vspace{-1cm}
\end{minipage}%
\hfill
\end{figure}

Finally, we study the effect of blocklength on the weighted sum rate performance in Fig.~\ref{fig11} and Fig.~\ref{fig12} for the cases of MRC and ZF, respectively. The rate targets for MRC and ZF are set as $R^{\rm{req}}=2$ bit/s/Hz and $R^{\rm{req}}=4$ bit/s/Hz, respectively. As expected, the system performance increases with the channel blocklength since  more time/frequency resource can be exploited for transmission. Some interesting observations can be found in these figures. First, when the blocklength is small, there is a significant performance gap between the proposed algorithm and the upper bound. However, this gap starts to reduce with the increase of the blocklength. This demonstrates that the blocklength has significant impact for IIoT devices with URLLC requirements. In addition, the proposed algorithm outperforms the  `Fixed pilot power alg.' algorithm  for both cases of MRC and ZF. This may be due to the fact that higher rate target is imposed in this example than that in Fig.~\ref{fig9}.

\section{Conclusions}\label{conclusion}
In this paper, we studied the resource allocation for uplink massive MIMO systems to support critical IIoT operating under finite channel blocklength, where multiple robots and/or actuators transmit URLLC signals to the central controller simultaneously. We first derived the closed-form data rate LB with imperfect CSI for both MRC and ZF receivers under the short packet transmission. We then formulate the weighted sum rate maximization to jointly optimize the pilot and payload power allocation while considering their energy, minimum data rate and decoding error probability requirements. This optimization problem is non-convex, and we proposed novel low-complexity iterative algorithms to solve it. Simulation results demonstrate that the algorithm converges rapidly, and outperforms the existing benchmark algorithms, especially the algorithm based on conventional Shannon capacity. This reveals the importance of adopting the achievable data expression for finite channel blocklength.

\begin{appendices}

\section{Proof of Theorem 1}\label{proofoftheorem1}

We first consider the MRC beamforming. The proof follows the similar steps as those in Appendix A of \cite{ngo2013energy} for perfect CSI. Denote $\gamma _k$ as the instantaneous SINR value when using MRC. By substituting ${{\bf{a}}_k} = {{{\hat{\bf h}}}_k}$ into  (\ref{moinou}), we have
\vspace{-0.2cm}
{\setlength\abovedisplayskip{3pt}
\setlength\belowdisplayskip{3pt}
\begin{equation}\label{dewfaef}
 \gamma _k = \frac{{p_k^d{{\left\| {{{{\hat{\bf h}}}_k}} \right\|}^4}}}{{\sum\nolimits_{i \in \mathcal{K}\setminus k} {p_i^d{{\left| {{\hat{\bf h}}_k^H{{{\hat{\bf h}}}_i}} \right|}^2}}  + \sum\nolimits_{i \in \mathcal{K}} {p_i^d{{\left| {{\hat{\bf h}}_k^H{{{\tilde{\bf h}}}_i}} \right|}^2}}  + {{\left\| {{{{\hat{\bf h}}}_k}} \right\|}^2}}}.
\end{equation}}
\!\!Then, ${{\mathbb{E}}\left\{ {\frac{1}{{{\gamma _k}}}} \right\}}$ can be expressed as
\vspace{-0.2cm}
{\setlength\abovedisplayskip{3pt}
\setlength\belowdisplayskip{3pt}
\begin{equation}\label{cfrahujk}
{\mathbb{E}}\left\{ {\frac{1}{{\gamma _k}}} \right\} ={\mathbb{E}} \left\{ {\frac{{\sum\nolimits_{i \in \mathcal{K}\setminus k} {p_i^d{{\left| {{u_{k,i}}} \right|}^2}}  + \sum\nolimits_{i \in \mathcal{K}} {p_i^d{{\left| {{v_{k,i}}} \right|}^2}}  + 1}}{{p_k^d{{\left\| {{{{\bf{\hat h}}}_k}} \right\|}^2}}}} \right\}
\end{equation}}
\!\!where ${u_{k,i}} = \frac{{{\hat{\bf h}}_k^H{{{\bf{\hat h}}}_i}}}{{{{\left\| {{{{\bf{\hat h}}}_k}} \right\|}}}}$ and ${v_{k,i}} = \frac{{{\hat{\bf h}}_k^H{{{\tilde{\bf h}}}_i}}}{{{{\left\| {{{{\hat{\bf h}}}_k}} \right\|}}}}$. Conditioned on ${\hat{\bf h}}_k$, ${u_{k,i}}$ and ${v_{k,i}}$ are Gaussian random variables with zero mean and variance equal to $\sigma _i$ and $\delta_i$, respectively. In addition, ${u_{k,i}}$ and ${v_{k,i}}$ are independent of ${{\hat{\bf h}}}_k$. Then, we have
\vspace{-0.2cm}
{\setlength\abovedisplayskip{3pt}
\setlength\belowdisplayskip{3pt}
\begin{equation}\label{nkjmyuj}
  {\mathbb{E}}\left\{ {\frac{1}{{\gamma _k}}} \right\} =\left( {\sum\nolimits_{i \in \mathcal{K}\setminus k} {p_i^d{\sigma _i}}  + \sum\nolimits_{i \in \mathcal{K}} {p_i^d{\delta _i}}  + 1} \right){\mathbb{E}}\left\{ {\frac{1}{{p_k^d{{\left\| {{{{\hat{\mathbf h}}}_k}} \right\|}^2}}}} \right\}.
\end{equation}}
\indent By using the identity \cite{tulino2004random}
{\setlength\abovedisplayskip{3pt}
\setlength\belowdisplayskip{3pt}
\begin{equation}\label{eagtrgtr}
\vspace{-0.2cm}
  {\mathbb{E}}\left\{ {{\rm{tr}}\left( {{{\bf{W}}^{ - 1}}} \right)} \right\} = \frac{m}{{n - m}}
\end{equation}}
\!\!where ${\bf{W}} \sim  {{\cal W}_m}\left( {n,{{\bf{I}}_n}} \right)$ is an $m \times m$ central complex Wishart matrix with $n$ ($n-m$) degrees of freedom. Then, we have
{\setlength\abovedisplayskip{3pt}
\setlength\belowdisplayskip{3pt}
\begin{equation}\label{iuheirty}
  {\mathbb{E}}\left\{ {\frac{1}{{p_k^d{{\left\| {{{{\hat{\mathbf h}}}_k}} \right\|}^2}}}} \right\}=\frac{1}{{p_k^d(M - 1){\sigma _k}}}, \  {\rm{for}}\  M\ge 2.
\end{equation}}
\indent By substituting (\ref{iuheirty}) into (\ref{nkjmyuj}), we can obtain ${{\hat \gamma }_k}$ in (\ref{jhoyjmkin}).

\section{Proof of Theorem 2}\label{proofoftheorem2}
By using ZF, we have ${{\bf{A}}^H}{\hat{\bf H}} = {{\bf{I}}_K}$. Thus, ${{\bf{a}}_k}{{{\hat{\bf h}}}_i}$ is equal to one if $k=i$; otherwise, it is equal to zero. Then, the instantaneous SINR for ZF can be rewritten as
{\setlength\abovedisplayskip{3pt}
\setlength\belowdisplayskip{-4pt}
\begin{equation}\label{anglbtbygt}
   \gamma _k = \frac{{p_k^d}}{{\sum\nolimits_{i \in \mathcal{K}} {p_i^d{{\left| {{\bf{a}}_k^H{{{\tilde{\bf h}}}_i}} \right|}^2}}  + {{\left\| {{{\bf{a}}_k}} \right\|}^2}}}.
\end{equation}}

Define  ${\bf{\Lambda}}  = {\rm{diag}}\left\{ {\sqrt {{\sigma_1}} , \cdots ,\sqrt {{\sigma_K}} } \right\}$ and ${\bf{\mathord{\buildrel{\lower3pt\hbox{$\scriptscriptstyle\smile$}}
\over H} }} = \left[ {\frac{{{{{\bf{\hat h}}}_1}}}{{\sqrt {{\sigma _1}} }}, \cdots ,\frac{{{{{\bf{\hat h}}}_K}}}{{\sqrt {{\sigma _K}} }}} \right]$. Then, we can express ${\bf{\hat H}}$ as ${\bf{\hat H}} = {\bf{\mathord{\buildrel{\lower3pt\hbox{$\scriptscriptstyle\smile$}}
\over H} \Lambda }}$. The columns of ${\bf{\mathord{\buildrel{\lower3pt\hbox{$\scriptscriptstyle\smile$}}
\over H} }}$ are independent of each other, and each column follows the distribution of $\mathcal{CN}(\mathbf{0}, \mathbf{I})$.
Then, ${{\mathbb{E}}\left\{ {\frac{1}{{{\gamma _k}}}} \right\}}$ can be expressed as
{\setlength\abovedisplayskip{3pt}
\setlength\belowdisplayskip{3pt}
\begin{eqnarray}
  {\mathbb{E}}\left\{ {\frac{1}{{\gamma _k}}} \right\} &=& \left. {{\left( {\sum\nolimits_{i \in \mathcal{K}} {p_i^d{\delta _i} + 1} } \right)\mathbb{E}\left\{ {{{\left[ {{{\left( {{{{\bf{\hat H}}}^H}{\bf{\hat H}}} \right)}^{ - 1}}} \right]}_{k,k}}} \right\}}} \middle/ {{p_k^d}} \right.\\
 &=& \left. {{\left( {\sum\nolimits_{i \in \mathcal{K}} {p_i^d{\delta _i} + 1} } \right){\mathbb{E}}\left\{ {{{\left[ {{{\left( {{\bf{\Lambda }}{{{\bf{\mathord{\buildrel{\lower3pt\hbox{$\scriptscriptstyle\smile$}}
\over H} }}}^H}{\bf{\mathord{\buildrel{\lower3pt\hbox{$\scriptscriptstyle\smile$}}
\over H} \Lambda }}} \right)}^{ - 1}}} \right]}_{k,k}}} \right\}}} \middle/ {{p_k^d}} \right.\\
 &=& \left. {{\left( {\sum\nolimits_{i \in \mathcal{K}} {p_i^d{\delta _i} + 1} } \right){\mathbb{E}}\left\{ {{{\left[ {{{\left( {{{{\bf{\mathord{\buildrel{\lower3pt\hbox{$\scriptscriptstyle\smile$}}
\over H} }}}^H}{\bf{\mathord{\buildrel{\lower3pt\hbox{$\scriptscriptstyle\smile$}}
\over H} }}} \right)}^{ - 1}}} \right]}_{k,k}}} \right\}}} \middle/ {{{\sigma _k}p_k^d}} \right.\\
&=& \left. {{\left( {\sum\nolimits_{i \in \mathcal{K}} {p_i^d{\delta _i} + 1} } \right){\mathbb{E}}\left\{ {{\rm{tr}}\left[ {{{\left( {{{{\bf{\mathord{\buildrel{\lower3pt\hbox{$\scriptscriptstyle\smile$}}
\over H} }}}^H}{\bf{\mathord{\buildrel{\lower3pt\hbox{$\scriptscriptstyle\smile$}}
\over H} }}} \right)}^{ - 1}}} \right]} \right\}}} \middle/ {{K{\sigma _k}p_k^d}} \right.\\
&=& \left. {{\left( {\sum\nolimits_{i \in \mathcal{K}} {p_i^d{\delta _i} + 1} } \right)}} \middle/ {{(M - K){\sigma _k}p_k^d}}\label{dewhfirhfr} \right.
\end{eqnarray}}
\!\!where (\ref{dewhfirhfr}) is obtained by using (\ref{eagtrgtr}).

\section{Proof of Lemma 2}\label{proofoflemma2}

For notational simplicity, we define function $T(x) \buildrel \Delta \over = 1 - \frac{1}{{{{(1 + {x})}^2}}}$. The second derivative of $T(x)$ w.r.t. $x$ is given by
\vspace{-0.2cm}
{\setlength\abovedisplayskip{3pt}
\setlength\belowdisplayskip{3pt}
\begin{equation}\label{hjyothj}
\vspace{-0.2cm}
  T''(x) = \frac{{ - 6}}{{{{\left( {{x} + 1} \right)}^4}}}\le 0.
\end{equation}}
\!\!Next, we prove that $G(x)$ is a concave function of $x$.  Since $T(x)$ is a concave function w.r.t. $x$, we have
\vspace{-0.3cm}
{\setlength\abovedisplayskip{3pt}
\setlength\belowdisplayskip{3pt}
\begin{equation}\label{uolyhik}
\vspace{-0.3cm}
  \theta T(\hat x) + (1 - \theta )T(\tilde x) \le T(\theta \hat x + (1 - \theta )\tilde x),
\end{equation}}
\!\!for $0\le \theta \le 1$, where $\hat x$ and $\tilde x$ are two different non-negative values. On the other hand, $\sqrt x $  is a concave function w.r.t. $x$. Then, for $0\le \theta \le 1$ we have
{\setlength\abovedisplayskip{3pt}
\setlength\belowdisplayskip{3pt}
\begin{equation}\label{dwfrf}
\vspace{-0.2cm}
  \theta \sqrt {T(\hat x)}  + (1 - \theta )\sqrt {T(\tilde x)}  \le \sqrt {\theta T(\hat x) + (1 - \theta )T(\tilde x)}.
\end{equation}}
\indent By combining (\ref{dwfrf}) with (\ref{uolyhik}), we have
{\setlength\abovedisplayskip{3pt}
\setlength\belowdisplayskip{3pt}
\begin{equation}\label{dXEWDEW}
 \theta \sqrt {T(\hat x)}  + (1 - \theta )\sqrt {T(\tilde x)}  \le \sqrt {T(\theta \hat x + (1 - \theta )\tilde x)},
\end{equation}}
\!\!which is equivalent to $\theta G(\hat x)  + (1 - \theta )G(\tilde x)  \le G(\theta \hat x + (1 - \theta )\tilde x)$. Hence, $G(x)$ is a concave function w.r.t. $x$, which completes the proof.

\section{Proof of Lemma 3}\label{proofoflemma3}

The equalities in (\ref{equalA}) can be readily proved by substituting the expressions of $\rho$ and $\eta$ in (\ref{joijojo}) and (\ref{swdefefr}) into (\ref{frejoig}).  Next, we focus on the proof of Inequality (\ref{frejoig}).

Define function $H(x) \buildrel \Delta \over = F(x) - G(x)$, and thus $H(\tilde x)=0$. Then, with some simple manipulations, the first-order derivative of function $H(x)$ w.r.t. $x$ can be calculated as
{\setlength\abovedisplayskip{4pt}
\setlength\belowdisplayskip{4pt}
\begin{equation}\label{wdweerghuj}
H'(x) = \frac{{\tilde x\sqrt {{x^2} + 2x} {{\left( {x + 1} \right)}^2} - x\sqrt {{{\tilde x}^2} + 2\tilde x} {{\left( {\tilde x + 1} \right)}^2}}}{{\sqrt {{{\tilde x}^2} + 2\tilde x} {{\left( {\tilde x + 1} \right)}^2}x\sqrt {{x^2} + 2x} {{\left( {x + 1} \right)}^2}}}.
\end{equation}}
\!\!Since both $x$ and $\tilde x$ are positive values, the sign of $H'(x)$ only depends on the nominator of $H'(x)$. Then, denote the nominator of $H'(x)$ as $J(x)$.

The function $J(x)$ can be rewritten as
{\setlength\abovedisplayskip{3pt}
\setlength\belowdisplayskip{3pt}
\begin{equation}\label{dwehiuhire}
 J(x) = x\tilde x\left( {\sqrt {1 + \frac{2}{x}} {{\left( {x + 1} \right)}^2} - \sqrt {1 + \frac{2}{{\tilde x}}} {{\left( {\tilde x + 1} \right)}^2}} \right).
\end{equation}}
\!\!Next, we show that $J(x)\le 0$ when $\frac{{\sqrt {17}  - 3}}{4} \le x \le \tilde x$, and $J(x)> 0$ when $ x > \tilde x$. In specific, let us define
{\setlength\abovedisplayskip{0pt}
\setlength\belowdisplayskip{0pt}
\vspace{-0.1cm}
\begin{equation}\label{dwejohoreuh}
\vspace{-0.1cm}
  U(x) \buildrel \Delta \over = \sqrt {1 + \frac{2}{x}} {\left( {x + 1} \right)^2}.
\end{equation}}
\!\!The first-order derivative of $U(x)$ w.r.t. $x$ is given by
{\setlength\abovedisplayskip{3pt}
\setlength\belowdisplayskip{3pt}
\begin{equation}\label{swdkoijhouhu}
  U'(x) = \frac{{x + 1}}{{{x^{1.5}}\sqrt {x + 2} }}\left( {2{x^2} + 3x - 1} \right).
\end{equation}}
\!\!It can be readily proved that $U'(x)\le 0$ when $0<x\le \frac{{\sqrt {17}  - 3}}{4}$, and $U'(x)>0$ when $x>\frac{{\sqrt {17}  - 3}}{4}$.
\\\indent As a result, when $\frac{{\sqrt {17}  - 3}}{4} \le x \le \tilde x$, we have $U(x)<U(\tilde x)$, and thus $J(x)<0$. From (\ref{wdweerghuj}), we know that $H'(x)<0$, which means $H(x)$ is a monotonically decreasing function of $x$. Hence, $H(x)\ge H(0)=0$ holds  for $\frac{{\sqrt {17}  - 3}}{4} \le x \le \tilde x$, which equivalently means $F(x)\ge G(x)$. On the other hand, when $x > \tilde x$,  we have $U(x)>U(\tilde x)$, and thus $J(x)>0$, which means $H'(x)>0$ and $H(x)$ is a monotonically increasing function of $x$. Hence, $H(x)>H(0)=0$ holds for $x\ge \tilde x$, and thus $F(x)>G(x)$.

Based on the above analysis, when $x\ge \frac{{\sqrt {17}  - 3}}{4}$, $F(x)$ is always no smaller than $G(x)$, which completes the proof.

\section{Proof of Theorem 3}\label{proofoftheorem3}

The first equation in  (\ref{jitiijyiij}) can be readily verified. Then, we focus on the second equality in (\ref{jitiijyiij}). The partial  derivative of $W({\breve{\bf x}})$ and $Y({\breve{\bf x}})$ w.r.t. $x_i$ are given by
{\setlength\abovedisplayskip{3pt}
\setlength\belowdisplayskip{3pt}
\begin{equation}\label{defr}
\frac{{\partial W({\bf{x}})}}{{\partial {x_i}}} = \prod\nolimits_{j \ne i} {(1 + {x_j})}, \frac{{\partial Y({\bf{x}})}}{{\partial {x_i}}} = \lambda {\tau _i}x_i^{ - 1}\prod\nolimits_{j\in \mathcal{K}} {x_j^{{\tau _j}}}, i=1,\cdots, K.
\end{equation}}
By substituting ${\bf{x}}=\breve{\bf x}$, $\lambda$ and ${\tau _i},\forall i$ into into the above two functions, we can show that
{\setlength\abovedisplayskip{3pt}
\setlength\belowdisplayskip{3pt}
\begin{equation}\label{dewe}
  {\left. {\frac{{\partial W({\bf{x}})}}{{\partial {x_i}}}} \right|_{{\bf{x}} = {\breve{\bf x}}}} = {\left. {\frac{{\partial Y({\bf{x}})}}{{\partial {x_i}}}} \right|_{{\bf{x}} = {\breve{\bf x}}}},i = 1, \cdots ,K.
\end{equation}}
Hence, the second equality in (\ref{jitiijyiij}) is proved.

Now, we begin to prove (\ref{jjorghot}). Before proceeding, we first introduce the following lemma.

\emph{\textbf{Lemma 5}}: For any given $\breve x\ge 0$, we have the following inequality:
{\setlength\abovedisplayskip{3pt}
\setlength\belowdisplayskip{3pt}
\begin{equation}\label{fjogjotji}
  \frac{{1 + x}}{{{x^\tau }}} \ge \frac{{1 + \breve x}}{{{{\breve x}^\tau }}},
\end{equation}}
where $\tau$ is given by $\tau  = \frac{{\breve x}}{{1 + \breve x}}$, and equality holds only when $x=\breve x$.

\emph{\textbf{Proof}}: \upshape  The proof is similar to those in Lemma 3, and thus omitted for simplicity. \hfill\rule{2.7mm}{2.7mm}

By applying inequality (\ref{fjogjotji}) for each $x_i,i=1,\cdots,K$, we have
{\setlength\abovedisplayskip{3pt}
\setlength\belowdisplayskip{3pt}
\begin{equation}\label{dwefrfre}
  \frac{{1 + {x_i}}}{{x_i^{{\tau _i}}}} \ge \frac{{1 + {{\breve x}_i}}}{{x_i^{{\tau _i}}}},i=1,\cdots,K.
\end{equation}}
Then, by multiplying the above $K$ inequalities, we have
{\setlength\abovedisplayskip{3pt}
\setlength\belowdisplayskip{3pt}
\begin{equation}\label{jiioihtr}
  \prod\nolimits_{i \in \mathcal{K}} {\frac{{1 + {x_i}}}{{x_i^{{\tau _i}}}}}  \ge \prod\nolimits_{i \in \mathcal{K}} {\frac{{1 + {x_i}}}{{x_i^{{\tau _i}}}}}
\end{equation}}
which completes the proof.

\end{appendices}

\bibliographystyle{IEEEtran}
\bibliography{myre}


\end{document}